\documentclass[]{aa}
\usepackage{subcaption}
\usepackage[varg]{txfonts}
\usepackage{natbib}
\usepackage{graphicx}

\makeatletter
\renewcommand*\aa@pageof{, page \thepage{} of \pageref*{LastPage}}
\makeatother

\bibpunct{(}{)}{;}{a}{}{,}
\usepackage[]{hyperref} 
\hypersetup{
    colorlinks,
    linkcolor=magenta,
    citecolor=blue,
    urlcolor=red}

\begin{document}

\title{The water-ice line as a birthplace of planets:\\
Implications of a species-dependent dust fragmentation threshold}

\author{Jonas M\"uller
\and Sofia Savvidou 
\and Bertram Bitsch}

\institute{Max-Planck-Institut für Astronomie, Königstuhl 17, 69117 Heidelberg, Germany}

%\date{Received <date> /Accepted <date>}

\keywords{Protoplanetary discs -- Planets and satellites: formation -- Hydrodynamics -- Opacity}

\abstract{The thermodynamic structure of protoplanetary discs is determined by dust opacities, which depend on the size of the dust grains and their chemical composition. In the inner regions, the grain sizes are regulated by the level of turbulence (e.g. $\alpha$ viscosity) and by the dust fragmentation velocity that represents the maximal velocity that grains can have at a collision before they fragment. Here, we perform self-consistently calculated 2D hydrodynamical simulations that consider a full grain size distribution of dust grains with a transition in the dust fragmentation velocity at the water-ice line.\ This approach accounts for the results of previous particle collision laboratory experiments, in which silicate particles typically have a lower dust fragmentation velocity than water-ice particles. Furthermore, we probe the effects of variations in the water abundance, the 
dust-to-gas ratio, and the turbulence parameter on the disc structure. For the discs with a transition in the dust fragmentation velocity at the water-ice line, we find a narrow but striking zone of planetary outward migration, including for low viscosities. In addition, we find a bump in the radial 
pressure gradient profile that tends to be located slightly inside the ice line. Both of these features are present for all tested disc parameters. Thus, we conclude that the ice line can function both as a migration 
trap, which can extend the growth times of planets before they migrate to the inner edge of the protoplanetary disc, and as a pressure trap, where planetesimal formation can be initiated or enhanced.}

\titlerunning{The water-ice line as a birthplace of planets: Implications of a species-dependent dust fragmentation threshold}
\authorrunning{J. M\"uller et al.}

\maketitle

%%%%%

\section{Introduction}

Planets are born in the protoplanetary disc that surrounds young stars for the first couple million years of their lifetimes. Dust particles in the disc grow to pebbles \citep{brauer2008} and planetesimals \citep[for a review, see][]{johansen2014}, causing the formation of planetary embryos. The growth of particles into pebbles was found to be facilitated at the 
water-ice line, where the local solid surface density is increased \citep{weidenschilling1977, hayashi1981} and water vapour condenses on grains \citep{ros2013}, preferably those that already have icy surfaces \citep{ros2019}, enabling them to grow to larger sizes than via pure coagulation. This increases 
the local solid-to-gas ratio and enhances planetesimal formation through the streaming instability \citep{ros2013, schoonenberg2017, drazkowska2017, yang2017}.
The ice line in itself has been suggested as the first site of planet formation due to the larger particles that can allow easier planetesimal, and thus planet, formation \citep{morbidelli2015, drazkowska2017, drazkowska2018}.

Pebbles interact with the gas disc and drift inwards \citep{weidenschilling1977, brauer2008}. The pebbles crossing the water-ice line lose their volatile contents due to evaporation and split into smaller grains \citep{schoonenberg2017etal}. The resulting pileup of small grains inside the water-ice line could also be a trigger of planetesimal formation \citep{ida2006, armitage2016}.

Since the position of the ice line is determined by the local disc temperature \citep{hayashi1981, sasselov2000, podolak2004}, the thermal structure of the disc is important for understanding planetesimal and planet 
formation. It is determined by dust opacities, which influence the magnitude of the heating and cooling processes in the disc and depend on the size of the dust grains and their chemical composition \citep{movshovitz2008, min2011}. Past simulations have mostly considered opacities from single grain sizes \citep{oka2011,bitsch2015,baillie2015}, with varying water abundances \citep{bitsch2016}. In order to calculate the position of the water-ice line as accurately as possible, it is essential to consider a full grain size distribution instead of a dust population with a single grain size \citep{savvidou2020}.

The grain size distribution is confined by a cutoff maximum grain size that itself depends on the thermal structure of the disc, the level of turbulence in the disc, and the dust fragmentation velocity \citep{brauer2008, birnstiel2011, birnstiel2012}, which is the relative velocity threshold 
at which fragmentation is always the outcome of a particle collision. Although there are contradictory findings \citep{musiolik2019}, laboratory experiments indicate that the dust fragmentation velocity of water-ice grains is larger than that of silicate particles \citep{poppe2000, gundlach2015}, which would result in a transition in the dust fragmentation velocity at the ice line. This would further drive apart the grain sizes and the 
dust opacities inside and outside the ice line and would greatly affect the structure of the disc.

Here we implement a transition in the dust fragmentation velocity at the ice line into the hydrodynamical code FARGOCA \citep{lega2014, bitsch2014, 
savvidou2020} and test variations in the composition of the dust, the solid-to-gas ratio, and the turbulence in the disc as well as how these factors affect the disc structure around the water-ice line. We then focus on the implications of our results on planetesimal formation and planet migration.

Our work is structured as follows. In Sect. \ref{methods} we explain the methods used to simulate the disc structure and describe our approach to modelling the transition in the dust fragmentation velocity. In Sect. \ref{results} we present our findings, and in Sect. \ref{discussion} we discuss the influences and implications of the disc structure on planetesimal and 
planet formation. Finally, we summarise in Sect. \ref{summary}.

%%%%%

\section{Methods} \label{methods}

\subsection{Hydrodynamical simulations}

The FARGOCA code solves the continuity and the Navier-Stokes equation and 
utilises the flux-limited diffusion (FLD) approach to radiative transfer. 
\citet{lega2014} and \citet{bitsch2014} introduced calculations with mean 
opacities derived from single grain sizes, as well as stellar irradiation, into the code. This was expanded by \citet{savvidou2020} by including grain size distributions, settling, and size- and composition-dependent opacities in order to account for a self-consistent treatment of the dust size distribution with respect to the thermal structure of the disc. More details on the hydrodynamical equations, opacities, and the implementation of the grain size distribution can be found in \citet{savvidou2020}.

The time evolution of the radiative energy density, $E_R$, and the thermal energy density, $\epsilon$, is determined by the following equations \citep{levermore1981,dobbs-dixon2010,commercon2011}:\begin{gather}
    \frac{\partial E_R}{\partial t}+\nabla\cdot\boldsymbol{\textbf{F}}=\rho\kappa_P[B(T)-cE_R]\\
    \frac{\partial\epsilon}{\partial t}+(\boldsymbol{\textbf{u}}\cdot\nabla)\epsilon=-P\nabla\cdot\boldsymbol{\textbf{u}}-\rho\kappa_P[B(T)-cE_R]+Q^++S.
\end{gather}
Here, $t$ is the time, $B(T)=4\sigma T^4$ is the black-body radiation (where $\sigma$ is the Stefan-Boltzmann constant), $c$ is the speed of light, $\rho$ is the gas volume density, $\kappa_P$ is the Planck opacity, $\boldsymbol{\textbf{u}}$ is the gas velocity, $P$ is the thermal gas pressure, $S$ is the stellar heating component, and $Q^+$ is the viscous dissipation or heating function that is computed in every grid cell. Since $Q^+$ is most efficient for high densities, viscous heating mostly originates from the midplane and is then radiatively diffused throughout the disc \citep[e.g.][]{kley2009}.

Using the FLD approach, the radiation flux is given by 
\citep{levermore1981}
\begin{gather}
    \boldsymbol{\textbf{F}}=-\frac{\lambda c}{\rho\kappa_R}\nabla E_R,
\end{gather}
where $\kappa_R$ is the Rosseland opacity and $\lambda$ denotes the flux limiter described in \citet{kley1989}. More details about the energy equations are available in \citet{bitsch2013}.

The stellar heating density received by a grid cell of width $\Delta r$ reads \citep{dobbs-dixon2010}
\begin{gather}
    S=F_*e^{-\tau}\frac{1-e^{-\rho\kappa_*\Delta r}}{\Delta r},
\end{gather}
where $F=R_*^2\sigma T_*^4/r^2$ is the stellar flux, $R_*$ is the stellar radius, $T_*$ is the stellar surface temperature, $\kappa_*$ is the stellar opacity, and $\tau$ is the radially integrated optical depth along the line of sight to the star up to each grid cell.

\subsection{Opacity-temperature module}

\begin{figure*}
    \centering
    \begin{subfigure}[l]{0.33\textwidth}
        \includegraphics[width=\textwidth]{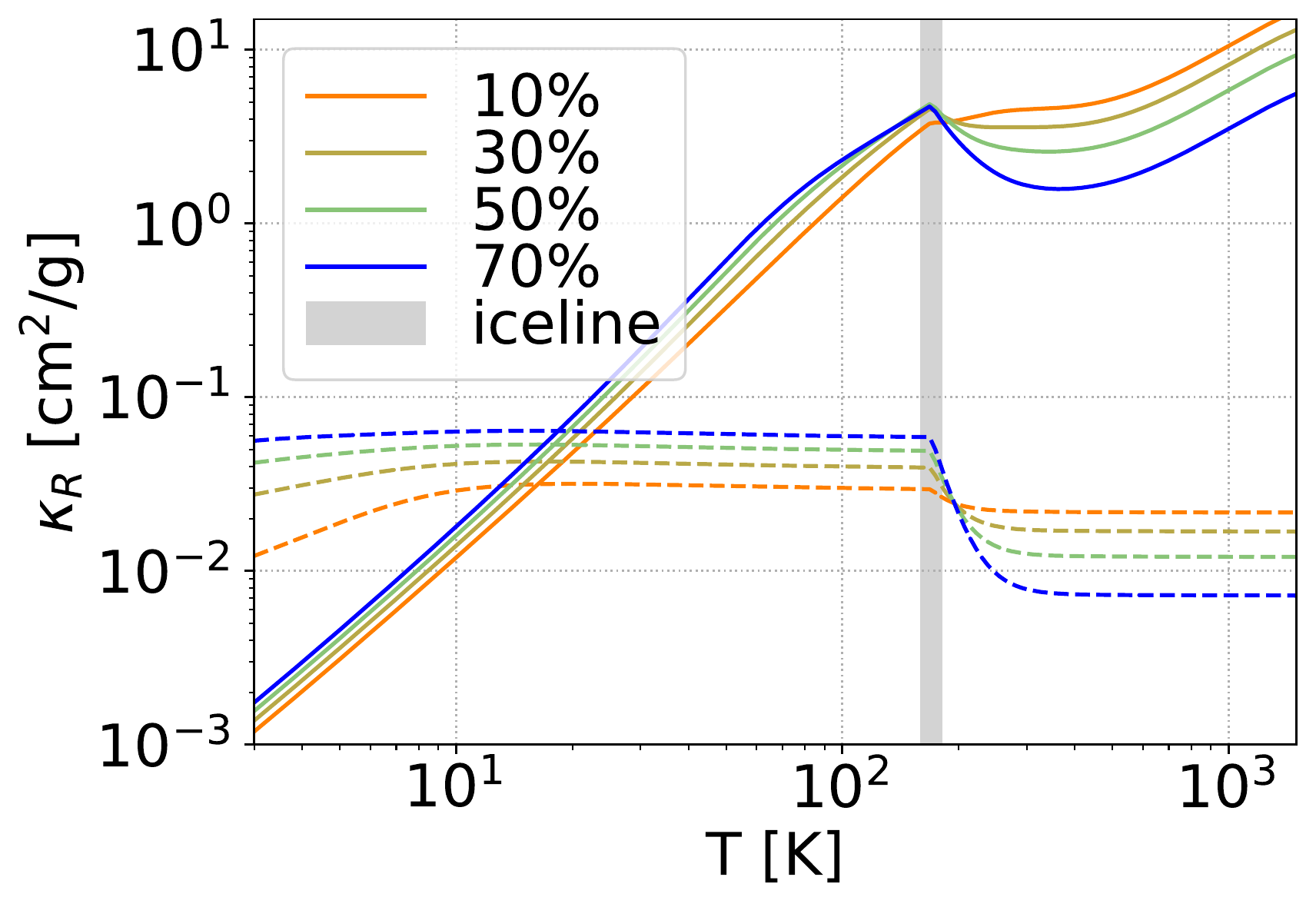}
    \end{subfigure}
    \begin{subfigure}[c]{0.33\textwidth}
        \includegraphics[width=\textwidth]{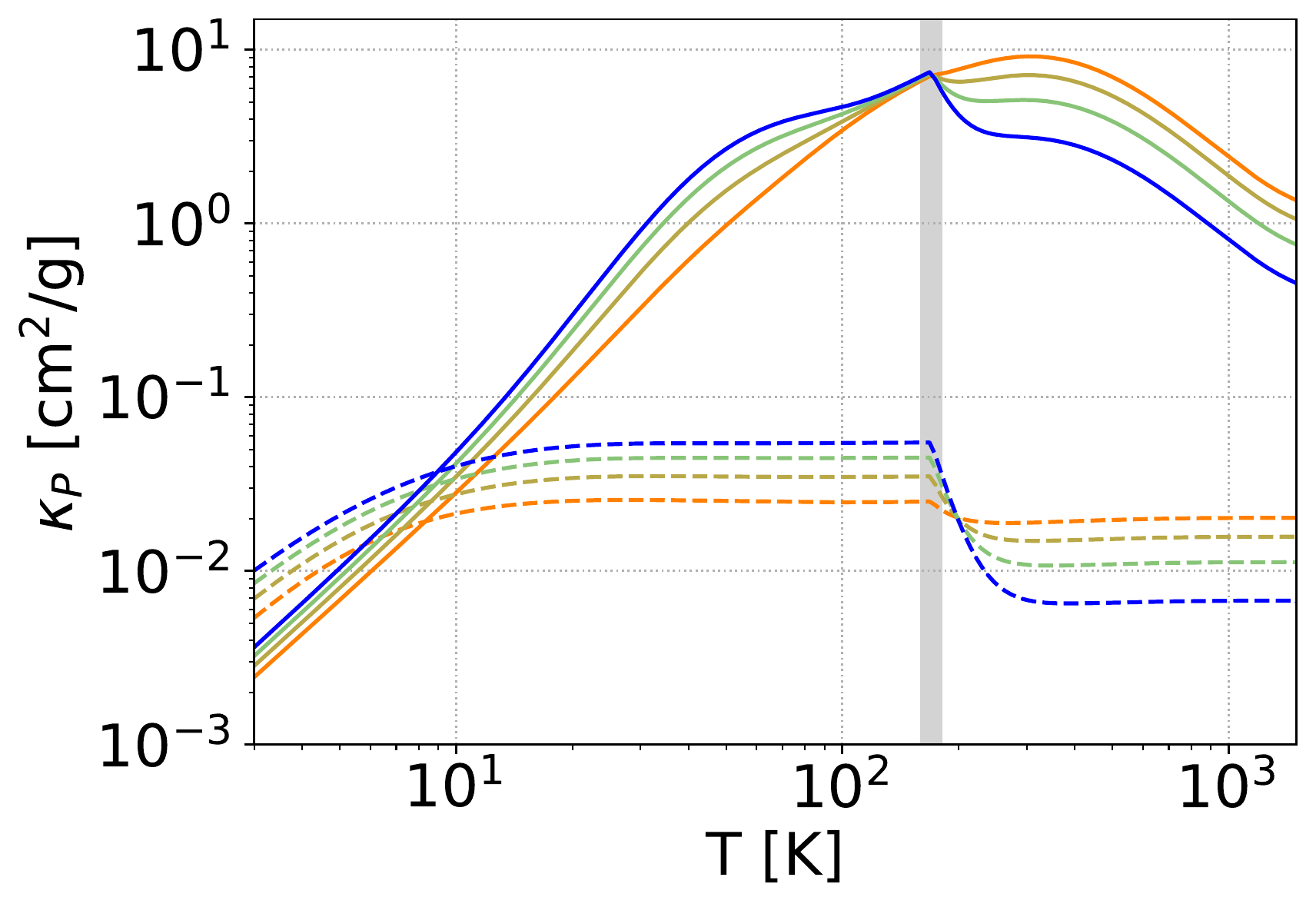}
    \end{subfigure}
    \begin{subfigure}[r]{0.33\textwidth}
        \includegraphics[width=\textwidth]{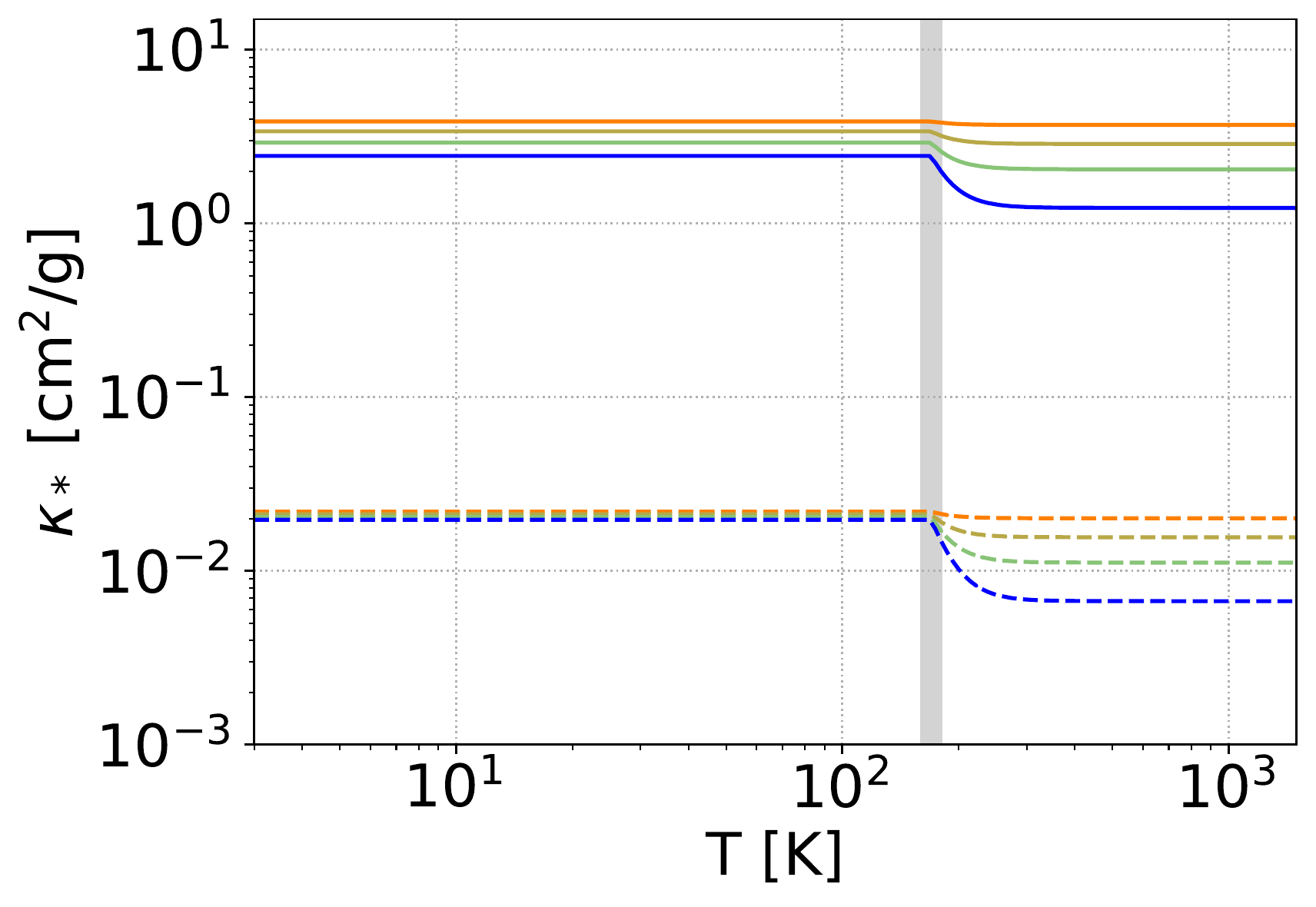}
    \end{subfigure}
    \caption{Rosseland, Planck, and stellar mean opacities (from left to right) as a function of temperature for grains with water abundances of 10\%, 30\%, 50\%, and 70\%. We calculated these opacities with RADMC-3D for a dust-to-gas ratio of 1\%. The solid lines correspond to a grain size of 1 $\mu$m, and the dashed lines correspond to a size of 1 mm. The water-ice line transition (vertical grey line) is located at (170 $\pm$ 10) K, causing a transition in opacity due to the evaporation and condensation of dust grains.}
    \label{fig: opacities}
\end{figure*}

For the opacities featured in the aforementioned equations, we utilised mean opacities that are averaged over all wavelengths. To compute the Planck, Rosseland, and stellar mean opacities as a function of the temperature, 
we used the RADMC-3D\footnote{\href{http://www.ita.uni-heidelberg.de/~dullemond/software/radmc-3d/}{http://www.ita.uni-heidelberg.de/$\sim$dullemond/software/radmc-3d/}}
code. The code calculates the wavelength-dependent opacities by utilising 
Mie-scattering theory and the optical constants for water ice \citep{warren2008} and silicates \citep{jaeger1994, dorschner1995}; it then averages them 
over all wavelengths. The main input parameters are the size of the grains, the dust-to-gas ratio of the disc, and the grain species. The opacity is always computed with a dust-to-gas ratio of 1\%, and we chose dust mixtures that are only made up of water ice and silicates, with water abundances ranging from 10\% to 70\% (a water abundance of e.g. 50\% 
corresponds to a water-to-silicate ratio of 1:1). Different water-ice ratios could result from different stellar abundances, where a large stellar C/O ratio could imply a lower water content \citep{bitsch2020}. When increasing the dust-to-gas ratio by a factor of two, the opacities shown in Fig. \ref{fig: opacities} also increase by a factor of two.

In other words, the total number of solids that are contained by the disc is fixed by the dust-to-gas ratio because the total mass of the gas disc 
does not change. The species of the solids is determined by the water abundance. Therefore, changing the water abundance does not influence the total mass of the dust in the disc.

In Fig. \ref{fig: opacities} we show the mean opacities as a function of the temperature for grains of 1 micron and 1 mm in size for four different water abundances. We only show two different grain sizes to illustrate the 
effects, but our code uses the full grain size distribution binned in at least ten opacity bins. Generally, the opacities of larger grains tend to be smaller and less temperature dependent, eventually resulting in colder 
discs.

At temperatures higher than that of the water-ice line transition, the water fraction of the grains sublimates and the opacities are only determined by the contribution of silicates. We did not consider gas opacities as they can be neglected at the temperatures reached in the discs in our simulations. The total dust mass was fixed by the dust-to-gas ratio, and thus a larger water fraction leads to a smaller amount of silicates being present, which in turn causes the opacities to be smaller at high temperatures for larger water abundances (Fig. \ref{fig: opacities}). At temperatures below the ice line transition, the deviations in the opacities for different water abundances stem from the optical constants of water-ice and silicate grains, leading to a larger Rosseland and Planck mean opacity for higher water fractions.

\subsection{Grain size distribution}

Collisions between two dust grains result in one of many possible outcomes depending on the relative velocity of the grains and their mass ratio \citep{weidenschilling1977, brauer2008, guettler2010}. The grain size distribution model we use in this work was inspired by the groundwork on dust distributions \citep{dohnanyi1969, mathis1977, tanaka1996} and only 
considers the equilibrium between fragmentation and coagulation. Although 
this is a simplification, \citet{savvidou2020} showed that the disc structure that emerges from this model is comparable to that resulting from the more complex model presented by \citet{birnstiel2011}, which also takes cratering and bouncing into 
account  \citep{zsom2010}, if the same cutoff maximum grain size is used. We picked the simpler model because it saves computational time. A review on grain growth is available in \citet{blum2008}.

The maximum grain size, $s_{max}$, of the distribution is defined as \citep{birnstiel2011}
\begin{gather}
    s_{max}\simeq\frac{2\Sigma_G}{\pi\alpha\rho_s}\frac{u_f^2}{c_s^2},
    \label{eq: max grain size}
\end{gather}

where $\alpha$ is the turbulence parameter of the $\alpha$ viscosity \citep{shakura1973}, $\Sigma_G$ is the gas surface density, $\rho_s=1.6$ g/cm$^2$ is the density of each particle, and $u_f$ is the dust fragmentation velocity, the threshold velocity at which fragmentation is always the outcome of a particle collision. The sound speed is given by
\begin{gather}
    c_s=\sqrt{\frac{k_BT}{\mu m_P}},
\end{gather}

where $k_B$ is the Boltzmann constant, $\mu=2.3$ is the mean molecular weight, and $m_P$ is the proton mass. The minimum grain size is set to 0.025 $\mu$m. We show the dust size distribution in Appendix \ref{further plots}.

\begin{figure}
    \resizebox{\hsize}{!}{\includegraphics{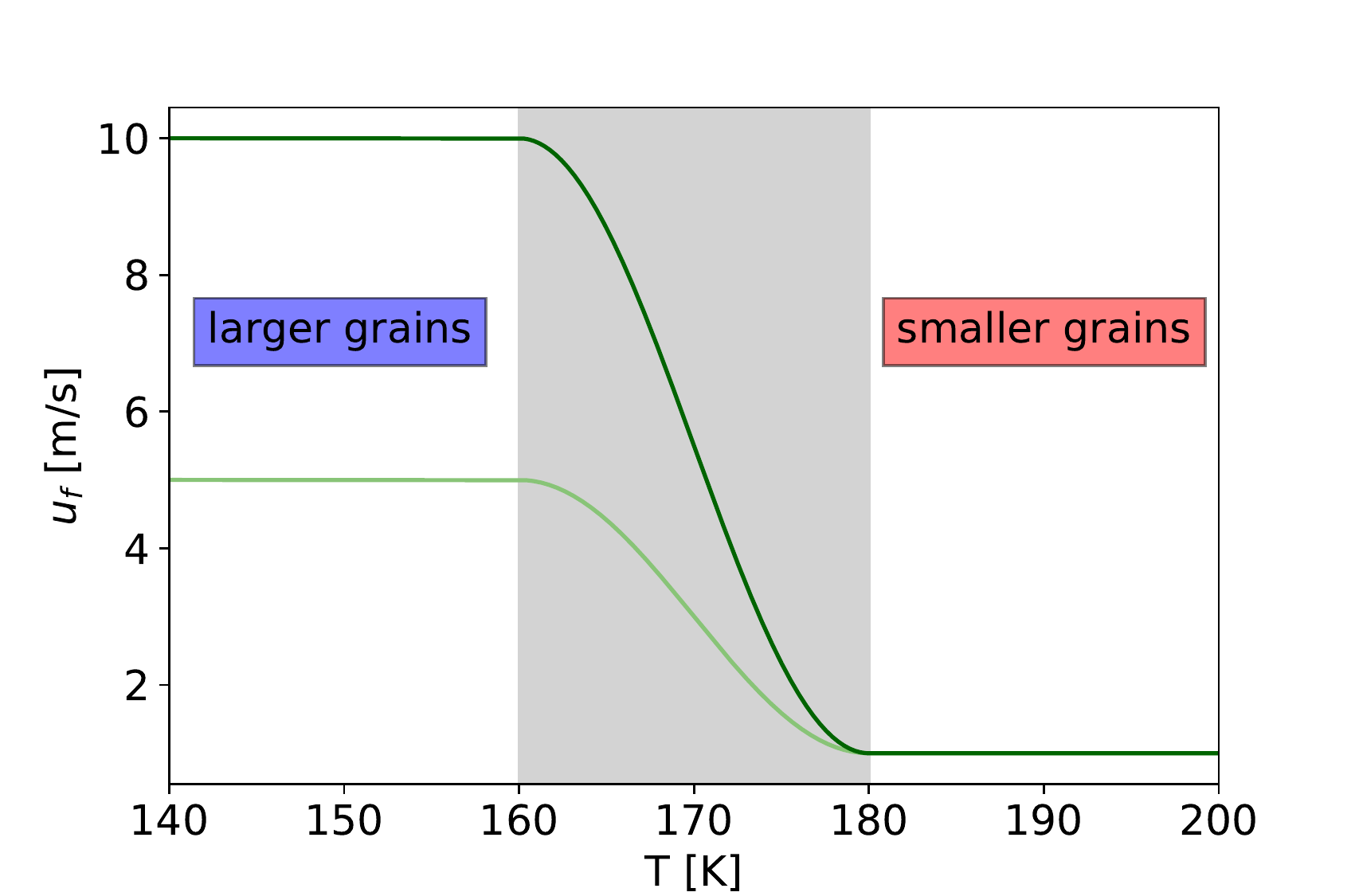}}
    \caption{Dust fragmentation velocity as a function of temperature. At 
temperatures $\geq 180$ K, the dust fragmentation velocity is 1 m/s, corresponding to the value that was found for silicate grains \citep{poppe2000}. In the icy regions we test larger values, as found for water-ice grains by \citet{gundlach2015}. We probe values of 5 m/s and 10 m/s. In 
the temperature range of the water-ice line transition (grey area), a simple sine connects the dust fragmentation velocity values.}
    \label{fig: uf transition}
\end{figure}

It has been found experimentally that silicate grains have a lower dust fragmentation velocity \citep[$u_f\sim 1$ m/s;][]{poppe2000} than water-ice grains \citep[$u_f\sim 10$ m/s;][]{gundlach2015}. However, recent experiments seem to indicate that the dust fragmentation velocity of water ice might be as low as 1 m/s \citep{musiolik2019}, though this has been called into question \citep{tatsuuma2019}. Here, we investigate how a transition 
in the dust fragmentation velocity at the water-ice line influences the thermal structure of the protoplanetary disc in order to account for the possibly different dust fragmentation velocity values of water-ice and silicate grains. The dust fragmentation velocity profiles we implement in this work are displayed in Fig. \ref{fig: uf transition}. When such a transition in the dust fragmentation velocity is utilised, the dust grains in the icy regions can grow to larger sizes than those in regions close to the star where only silicate grains are present, because they have a lower dust fragmentation velocity (Eq. \ref{eq: max grain size}).

It should be noted that the effects of radial grain drift \citep{weidenschilling1977, brauer2008} are not considered by the code. However, in our simulations the fragmentation size limit is always lower than the radial drift limit, and thus the grains have already fragmented into smaller pieces before they can experience a significant amount of radial drift. Therefore, drift can be neglected for the parameters chosen in our simulations. We discuss the implications of radial drift on our results further in Sect. \ref{discussion}.

\subsection{Simulation setups}

In our simulations the stellar mass is $M_*=1M_\odot$, the stellar radius is $R_*=1.5R_\odot$, and the surface temperature of the star is 4370 
K. We only considered equilibrium discs, meaning that the total mass of the 
disc is fixed in time and there is no accretion of material onto the star. The initial gas surface density profile reads
\begin{gather}
    \Sigma_G=\Sigma_{G,0}\cdot(r/\text{AU})^{-1/2},
\end{gather}
and the initial gas surface density, $\Sigma_{G,0}$, is 1000 g/cm$^2$. We only simulated the upper half of the protoplanetary disc and assumed symmetry towards the lower half in order to save computation time. The temperature of the interstellar medium is set to 3 K, so the disc is always cooled 
by its upper boundary \citep[see][]{bitsch2013}. The viscosity follows an 
$\alpha$ prescription and is given by 
\begin{gather}
    \nu=\alpha\frac{c_s^2}{\Omega_K}=\alpha c_s H,
\end{gather}
where $\Omega_K$ denotes the Keplerian orbit frequency and $H$ the scale height of the disc.

In this work, we present three sets of simulations, one for each of the $\alpha$ values of $5\times10^{-3}$, $10^{-3}$, and $5\times10^{-4}$. A set consists of three individual simulations that differ from one another only through their dust fragmentation velocity profile: Two of them feature a constant, temperature-independent dust fragmentation velocity with the values 1 m/s and 5 m/s. The third simulation features a transition in the dust fragmentation velocity from 1 m/s to 5 m/s, as displayed in Fig. \ref{fig: uf transition}. In addition, we tested for the highest viscosity value used in this work, $\alpha = 5\times 10^{-3}$, as well as the 
influence of different water abundances and dust-to-gas ratios and the influence of a larger dust fragmentation velocity, 10 m/s.

\begin{figure*}
    \centering
    \begin{subfigure}{0.42\textwidth}
        \includegraphics[width=\textwidth]{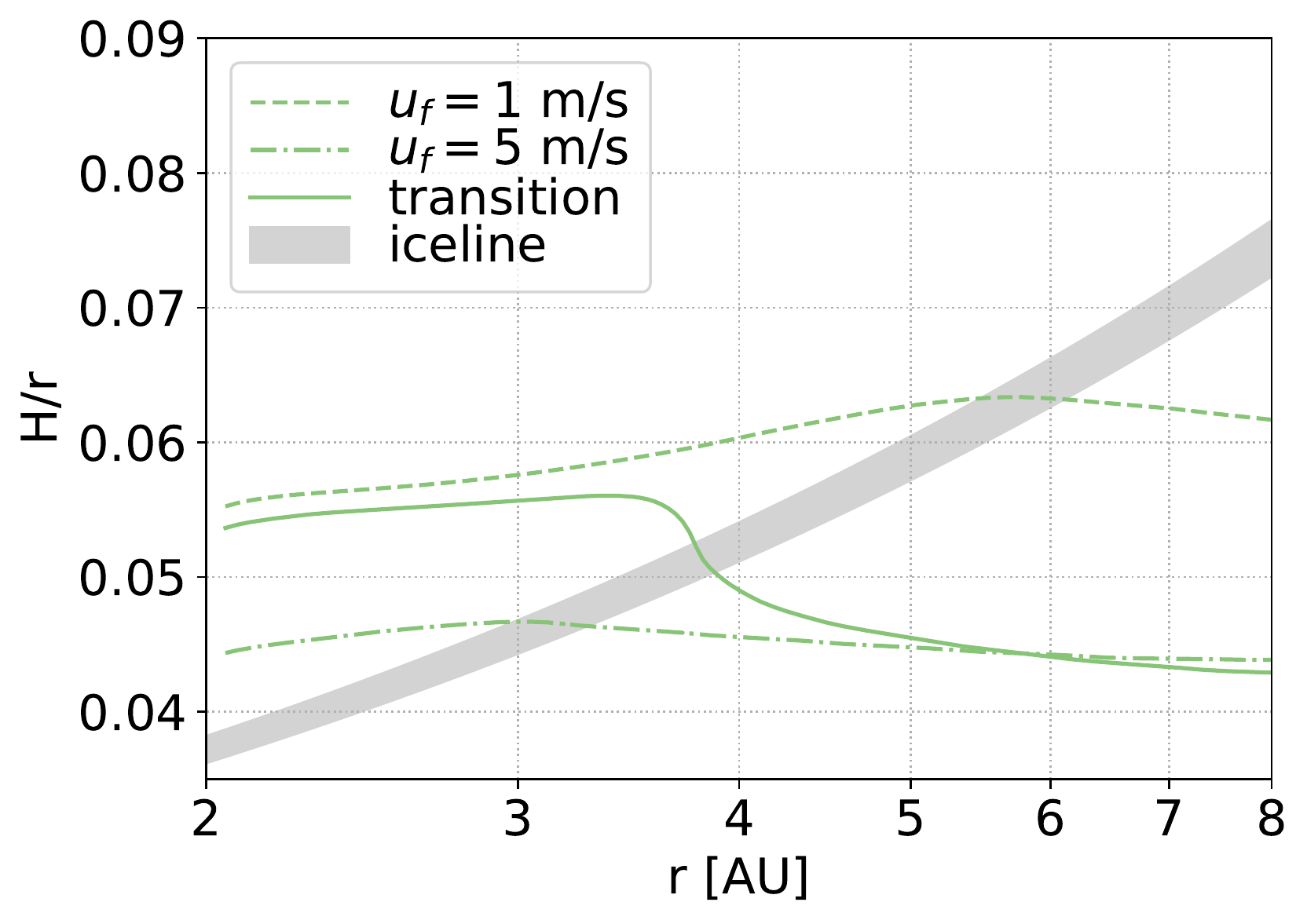}
    \end{subfigure}
    \begin{subfigure}{0.42\textwidth}
        \includegraphics[width=\textwidth]{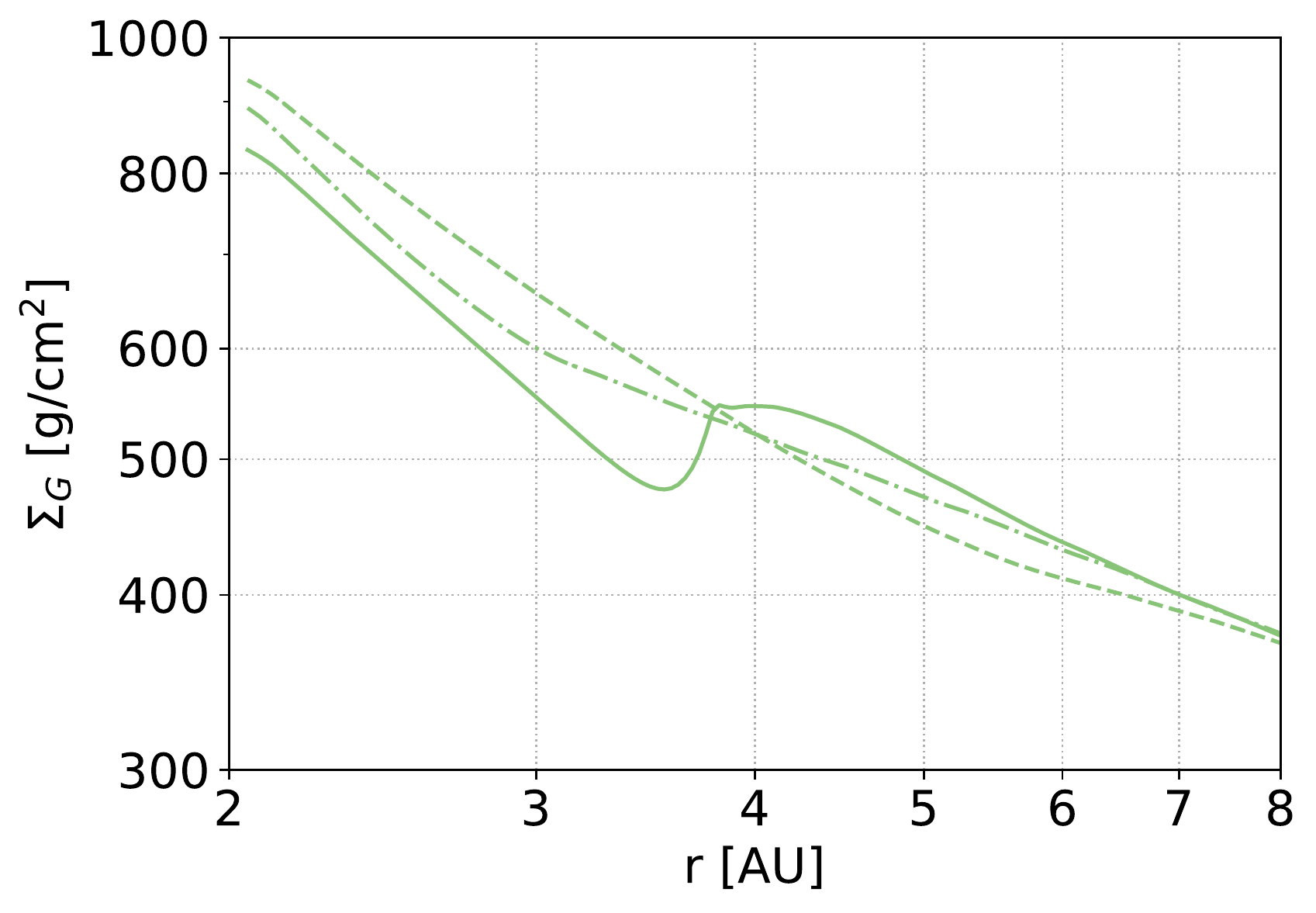}
    \end{subfigure}
    \caption{Aspect ratio (left) and gas surface density (right) as a function of the orbital distance from the star for different dust fragmentation velocity profiles. The dashed and dash-dotted lines mark the discs with a constant dust fragmentation velocity of 1 m/s and 5 m/s, respectively, while the solid line indicates the disc that features a transition in the dust fragmentation velocity between those values (Fig. \ref{fig: uf transition}). The water abundance is 50\%, the dust-to-gas ratio is 1\%, and the turbulence parameter is $\alpha=5\times10^{-3}$. We only show the region close to the water-ice line, which is marked by the grey band.}
    \label{fig: high viscosity}
\end{figure*}

\begin{figure*}
    \centering
    \begin{subfigure}{0.42\textwidth}
        \includegraphics[width=\textwidth]{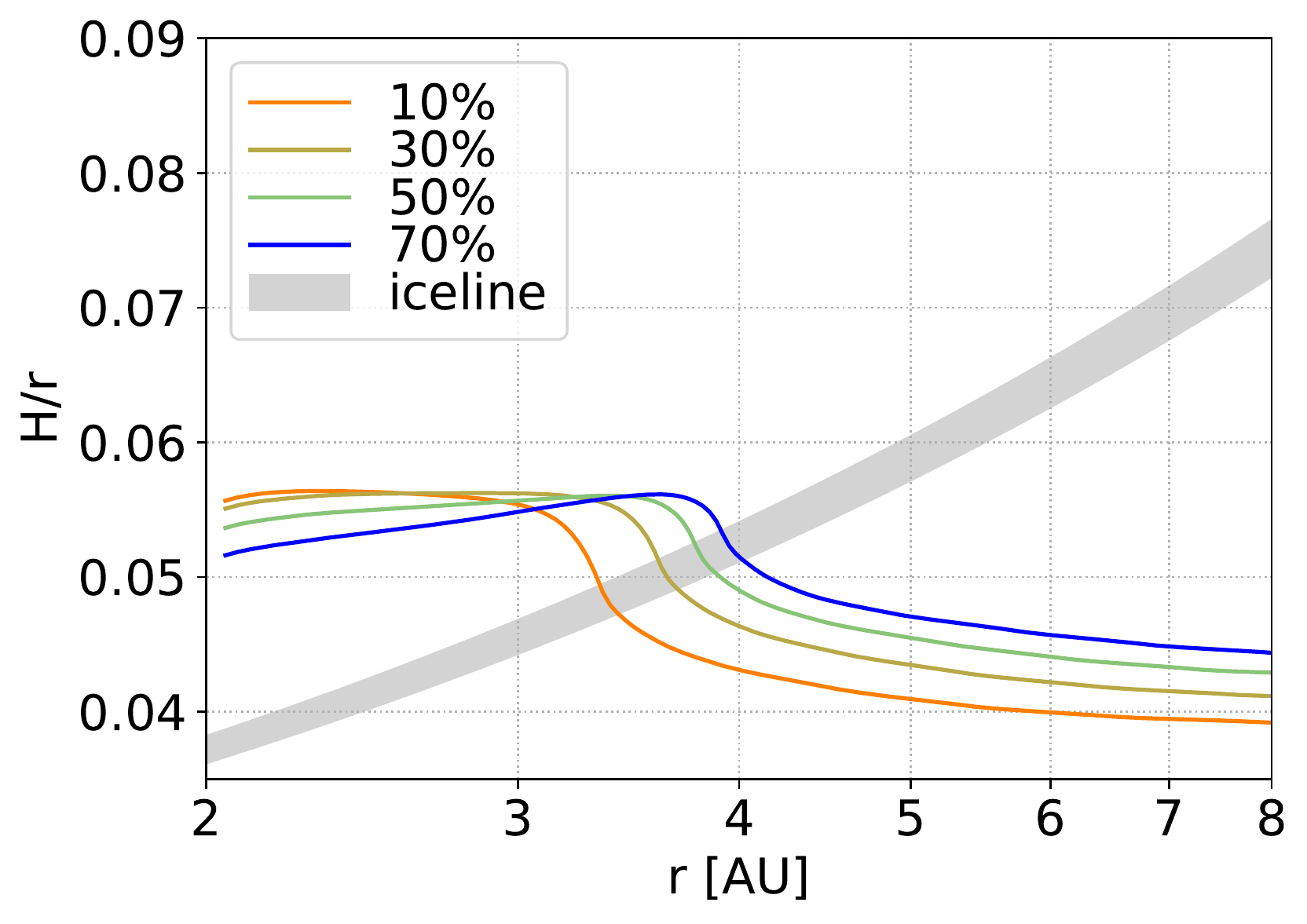}
        \caption{}
        \label{subfig: water to silicate ratio AR}
    \end{subfigure}
    \begin{subfigure}{0.42\textwidth}
        \includegraphics[width=\textwidth]{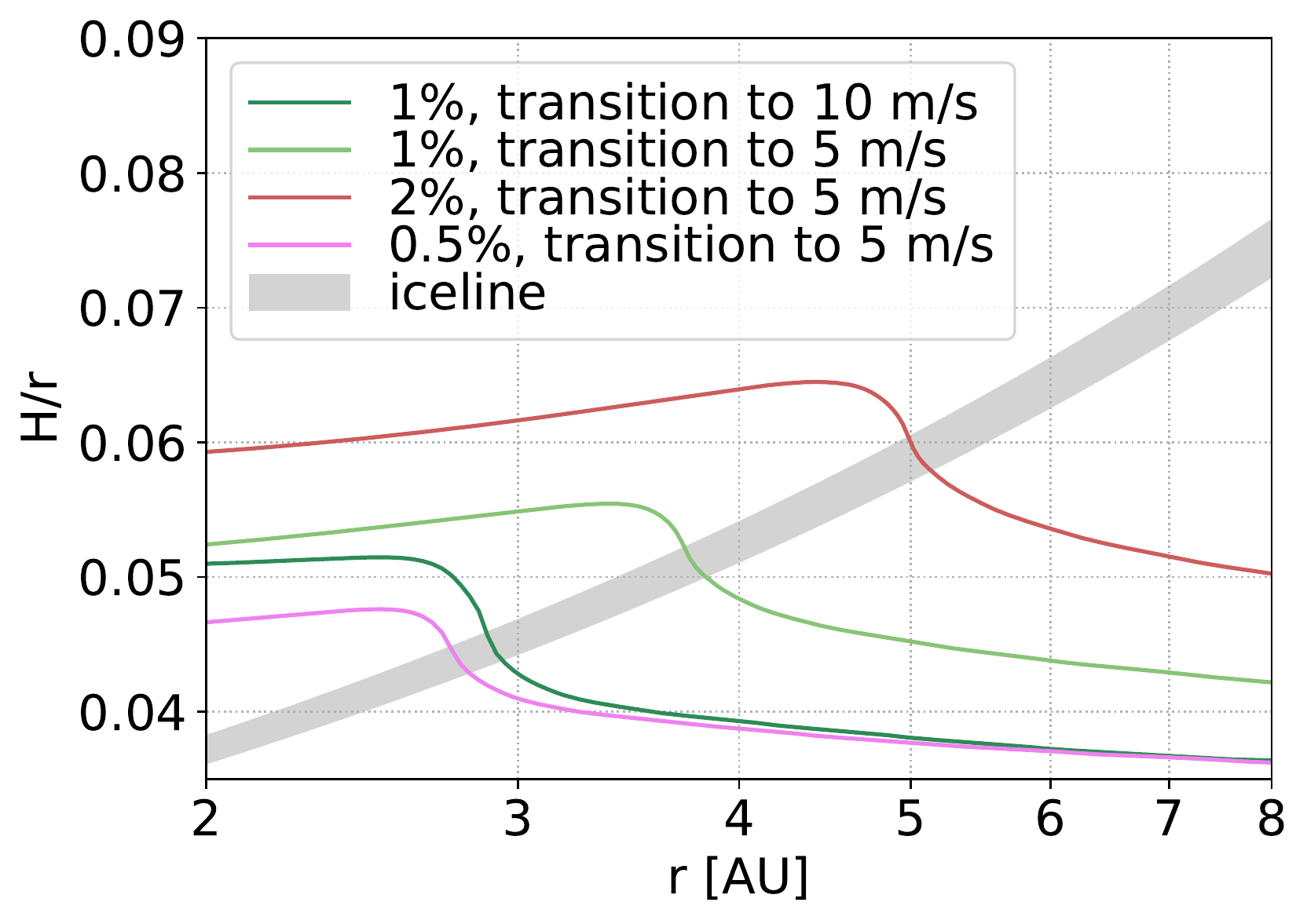}
        \caption{}
        \label{subfig: fragmentation velocity AR}
    \end{subfigure}
    \caption{Aspect ratio as a function of the orbital distance for discs with a viscosity of $\alpha=5\times10^{-3}$.
    {\bf Left:} Aspect ratio for discs with different water abundances that feature a transition in the dust fragmentation velocity from 1 m/s to 5 m/s and a dust-to-gas ratio 
of 1\%. {\bf Right:} 
    Aspect ratio of discs with different dust-to-gas ratios that feature a transition in the dust fragmentation velocity from 1 m/s to 5 m/s. Furthermore, we show the effects of a 
transition in the fragmentation velocity from 1 m/s to 10 m/s for a disc with a dust-to-gas ratio of 1\%.}
\end{figure*}

The simulations are embedded in four different simulation boxes that vary 
in terms of the radial distance covered by the discs, the opening angle, and the number of radial and vertical grid cells. We used the simulation boxes that 
only extend to smaller orbital distances and have smaller disc opening angles for colder discs because larger boxes lead to many optically thin grid cells above the real disc, which significantly increases the computation time. In the present work, we focus on the water-ice line transition region that is covered by the different simulation boxes well.

For the simulations featuring $\alpha=5\times10^{-3}$, a dust-to-gas ratio of 1\%, and a dust fragmentation velocity of 1 m/s and/or 5 m/s, we used 1440$\times$70 (radial-vertical direction) grid cells and an opening angle of 18$^{\circ}$, and the disc extends from 2 to 50 AU (total mass of the gas disc: $\sim0.08524\ M_\odot$). We also used this setup for simulations with varying water abundance.

The second set of simulations featured 285$\times$70 grid cells and used an opening angle of $14^{\circ}$. The discs simulated in this box extend from 0.5 to 10 AU (total mass: $\sim0.00760\ M_\odot$) and feature $\alpha=5\times10^{-3}$ and a fixed water abundance of 50\%. Using this environment, we executed a simulation with a transition in the dust fragmentation velocity to 10 m/s and investigated variations in the dust-to-gas ratio for 
a disc that features a transition in the dust fragmentation velocity to 5 
m/s.

The simulations with $\alpha=10^{-3}$ and $5\times10^{-4}$ used 147$\times$40 grid cells with a disc extending from 0.1 to 5 AU. The opening angle was $7^{\circ}$ for the discs featuring $\alpha=10^{-3}$ (total mass: $\sim0.00270\ M_\odot$) and $6^{\circ}$ for the discs featuring $\alpha=5\times10^{-4}$ (total mass: $\sim0.00267\ M_\odot$). For the discs featuring a transition in the dust fragmentation velocity, we increased the number of radial grid cells to 490 (i.e. $\Delta r = 0.01$ AU) for the last 300 orbits to properly resolve the transition at the ice line. The influence of the radial resolution on the transition region is further discussed in Appendix \ref{sec: resolution}. All other simulations featured the same radial resolution ($\Delta r = 0.03333$ AU).
The simulations presented in the following ran until they reached thermal 
equilibrium.

%%%%%

\section{Results} \label{results}

\begin{figure*}[ht]
    \centering
    \begin{subfigure}{0.42\textwidth}
        \includegraphics[width=\textwidth]{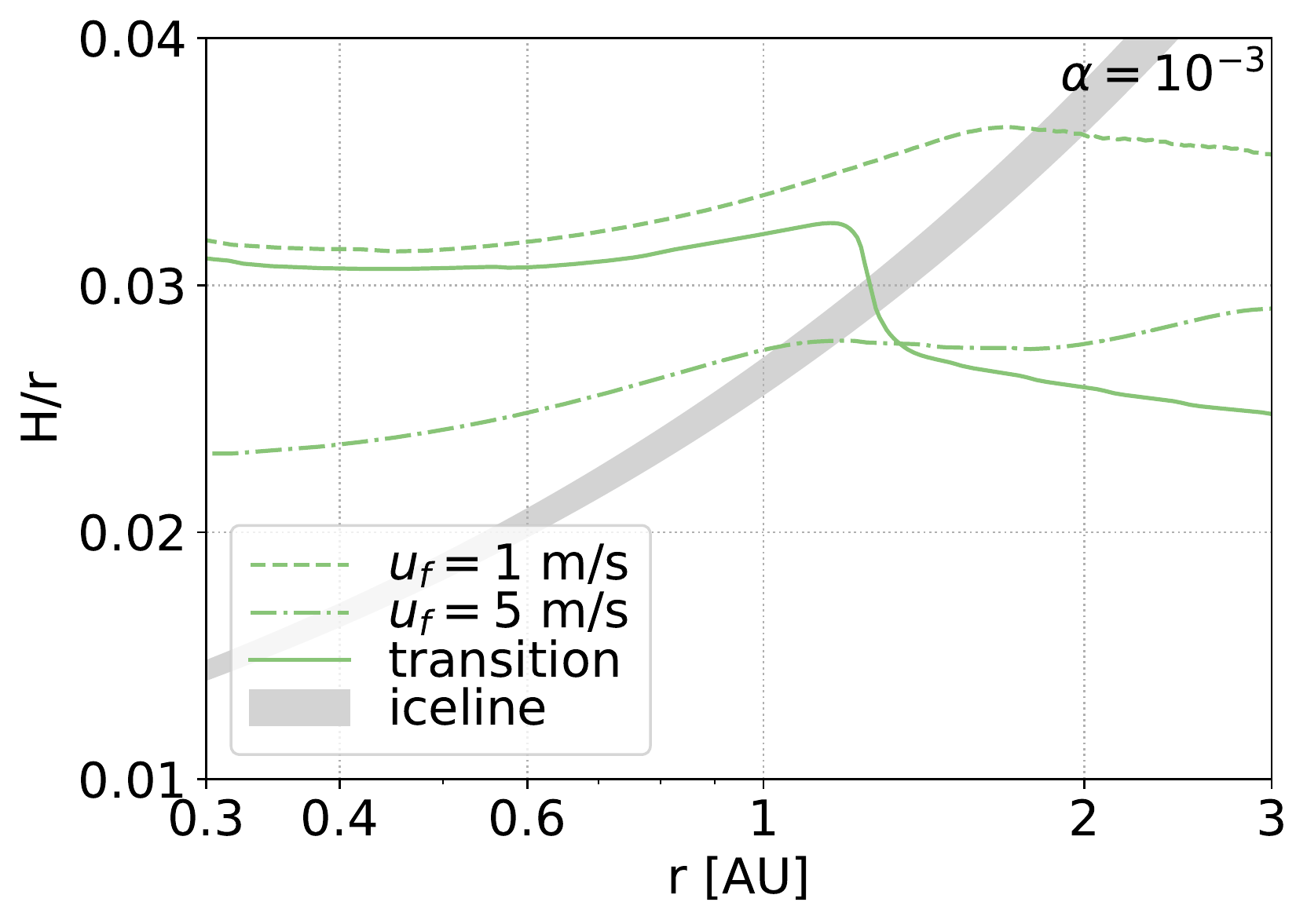}
    \end{subfigure}
    \begin{subfigure}{0.42\textwidth}
        \includegraphics[width=\textwidth]{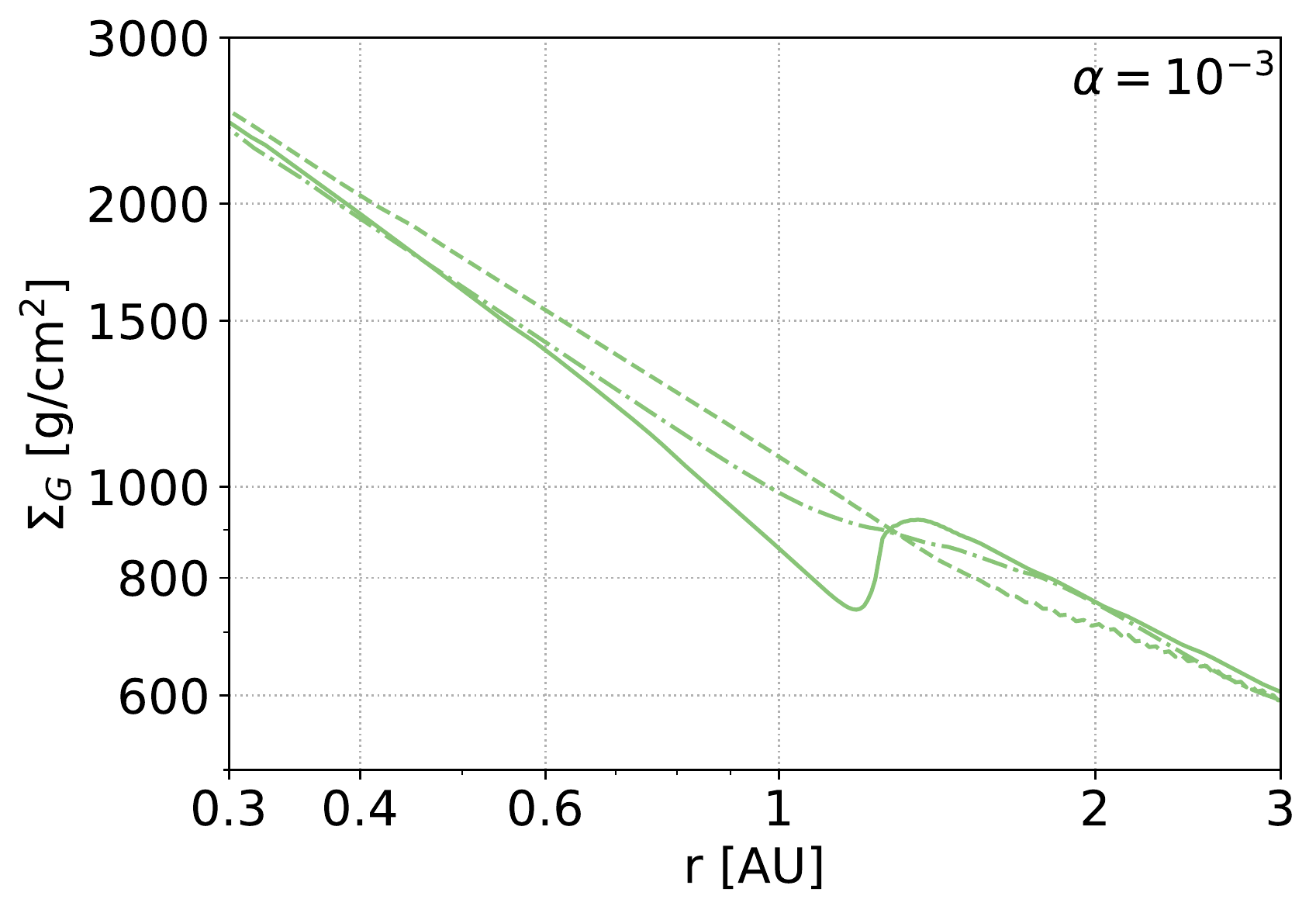}
    \end{subfigure}
    \begin{subfigure}{0.42\textwidth}
        \includegraphics[width=\textwidth]{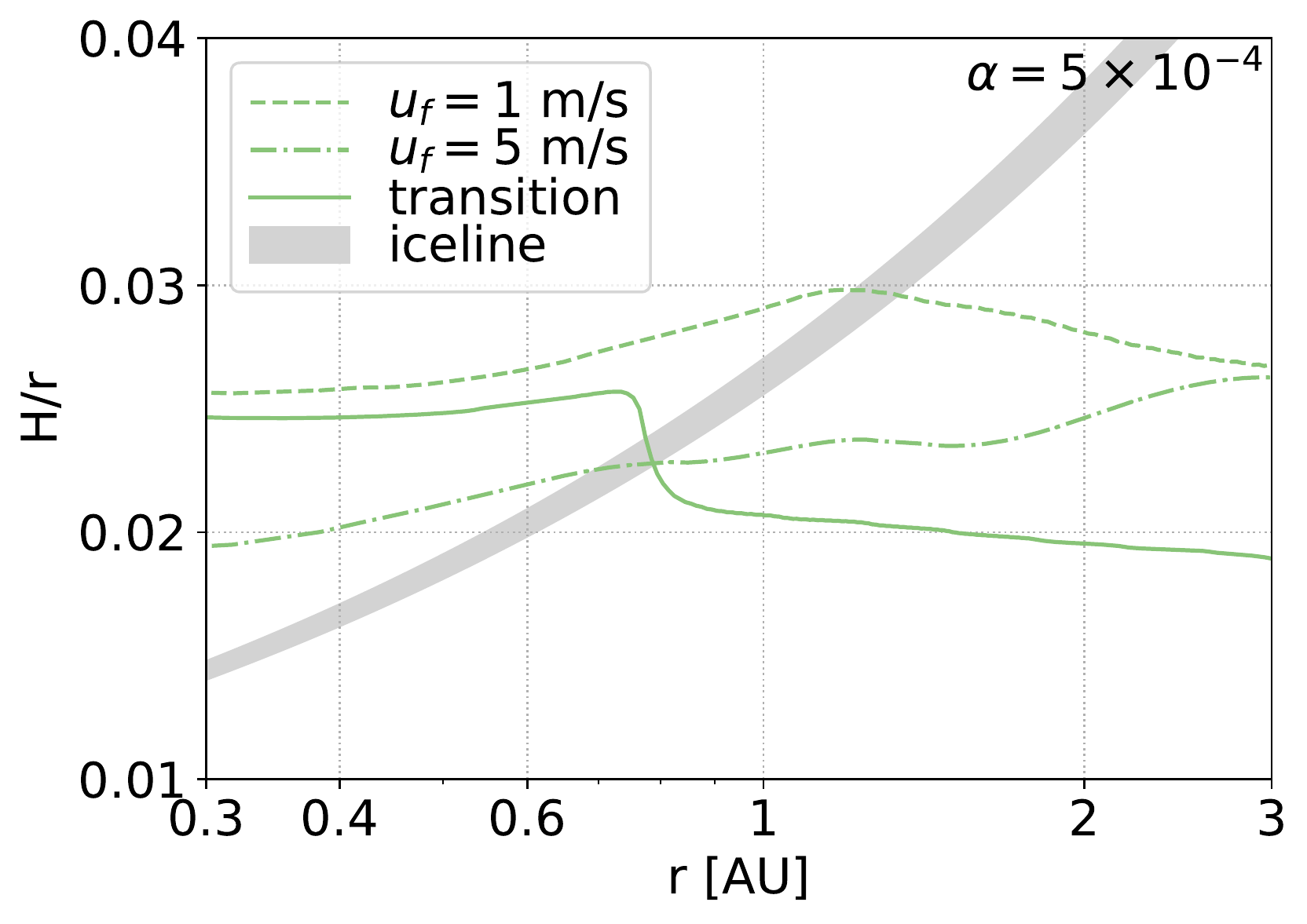}
    \end{subfigure}
    \begin{subfigure}{0.42\textwidth}
        \includegraphics[width=\textwidth]{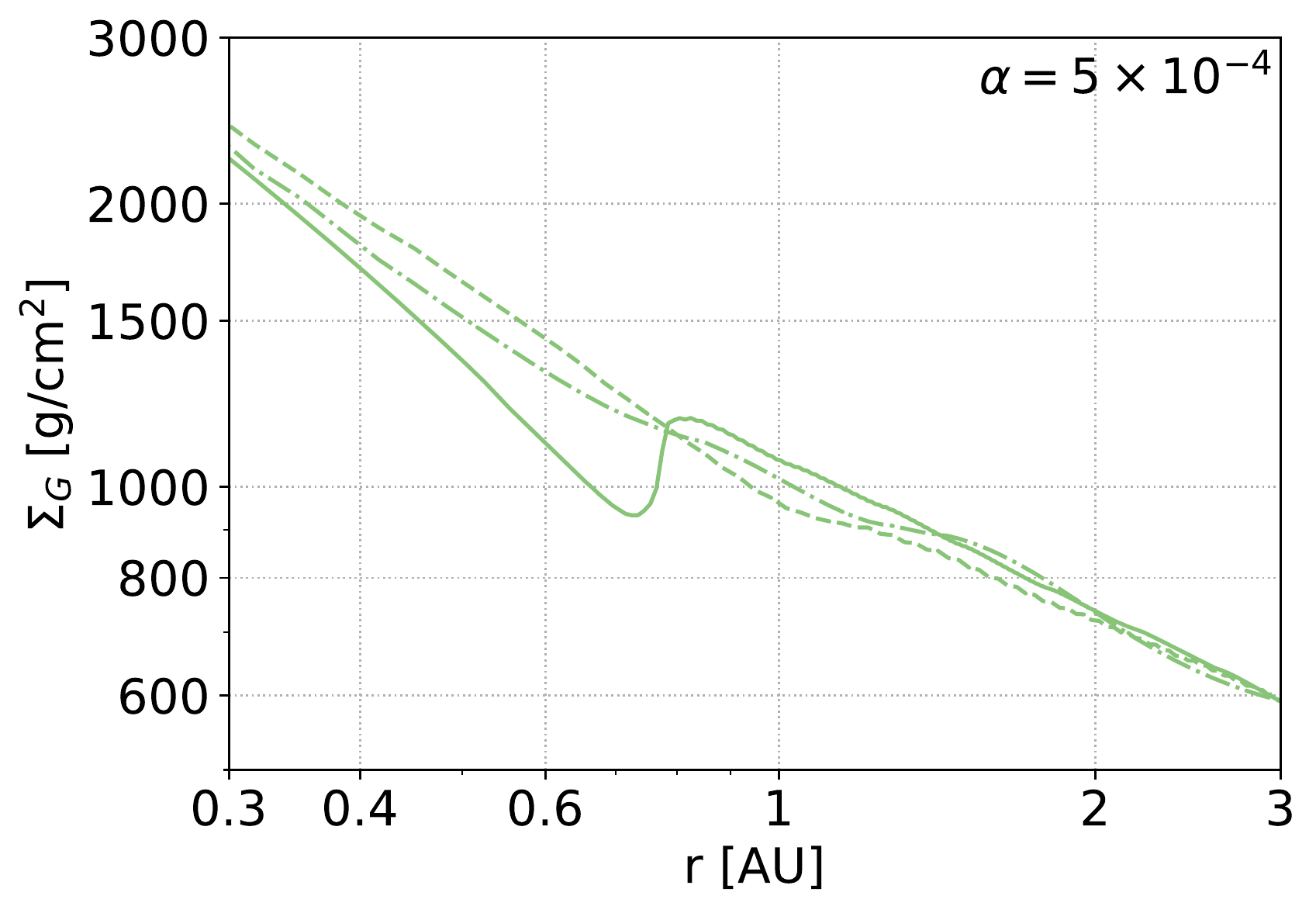}
    \end{subfigure}
    \caption{Aspect ratio (left column) and gas surface density (right column) as a function of the orbital distance for discs with a viscosity of 
$\alpha=10^{-3}$ (upper row) and $\alpha=5\times10^{-4}$ (lower row) using a water abundance of 50\%, a dust-to-gas ratio of 1\%, and different dust fragmentation velocity profiles.}
    \label{fig: low viscosity}
\end{figure*}

\subsection{High viscosity discs}

In this section we compare the structure of the discs that emerge from a constant dust fragmentation velocity to discs that harbour a transition in the dust fragmentation velocity for a high viscosity ($\alpha=5\times10^{-3}$). We tested variations in the water abundance and the dust-to-gas ratio as well as different dust fragmentation velocity values.

At first, we chose a water abundance of 50\% and a dust-to-gas ratio of 1\% and compared the disc featuring a transition in the dust fragmentation velocity from 1 m/s to 5 m/s (Fig. \ref{fig: uf transition}) to discs that have a constant dust fragmentation velocity profile. The resulting disc 
structures can be seen in Fig. \ref{fig: high viscosity}. The grey area indicates the location of the ice line at $(170\pm10)$ K. In this area, the aspect ratios of the discs with a constant dust fragmentation velocity both feature a bump that is induced by the change in opacity at 170 K (Fig. \ref{fig: opacities}). At these high opacities, this bump is in a region where viscous heating dominates \citep{bitsch2015}. In addition, the aspect ratio of the disc where the dust grains have a constant fragmentation velocity of 5 m/s is smaller than the aspect ratio of the disc where the dust fragmentation velocity is set to 1 m/s. This is due to the dust grains being able to grow to larger sizes for higher dust fragmentation velocity values (Eq. \ref{eq: max grain size}), leading to colder discs, in which the ice line is located at smaller orbital distances from the star.

Contrary to the disc profiles that feature a constant dust fragmentation velocity, a transition of the dust fragmentation velocity results in a sharp 
decrease in the aspect ratio at the ice line. In this case, the aspect ratio is comparable to that of the disc where grains have a constant dust fragmentation velocity of 1 m/s inside the ice line as well as to that of a disc where the dust fragmentation velocity is set to 5 m/s outside the ice line.

Figure \ref{fig: high viscosity} displays the gas surface density as a function of the orbital distance for the same set of discs. The surface densities of the discs where the grains have a constant fragmentation velocity 
only show small deviations. The disc with a transition in the dust fragmentation velocity features a distinct increase in the gas surface density near the ice line. This rearrangement of the gas surface density is caused 
by the changes in the disc's aspect ratio and temperature, which ultimately change the viscosity, in turn influencing the transport of gas in the disc.

Next, we investigated the effects of changes in the water abundance, the dust-to-gas ratio, and the dust fragmentation velocity on the thermal structure of discs that feature a transition in the dust fragmentation velocity. 
In Fig. \ref{subfig: water to silicate ratio AR} we show the aspect ratio 
as a function of the orbital distance for discs with water abundances of 10\%, 30\%, 50\%, and 70\%. The dust-to-gas ratio is set to 1\%. We can see that the aspect ratio is smaller for a lower water abundance outside the ice line, causing the ice line to be located closer to the star. This deviation is due to the different opacities of the dust mixtures (Fig. \ref{fig: opacities}) and was also found in hydrodynamical simulations with single grain sizes \citep{bitsch2016}. Apart from that, the thermal structure of discs with different water contents are comparable, and all feature a significant change in the aspect ratio at the water-ice line.

In Fig. \ref{subfig: fragmentation velocity AR} we compare the aspect ratio of a disc with a transition in the dust fragmentation velocity from 1 m/s to 5 m/s to a disc with a transition from 1 m/s to 10 m/s. In addition, we display the aspect ratio of discs that feature a transition in the dust fragmentation velocity to 5 m/s for dust-to-gas ratio values of 0.5\%, 1\%, and 2\%. All discs shown in Fig. \ref{subfig: fragmentation velocity 
AR} contain a water fraction of 50\%.

Since the dust fragmentation velocity is larger outside the ice line for the transition to 10 m/s, the dust grains can grow to larger sizes in that region, causing the opacity of the disc, and thus the aspect ratio, to be smaller. Consequently, the disc cools more efficiently and the ice line is located closer to the star. In addition, the difference in the aspect ratio inside and outside the ice line is larger, making the transition even more pronounced.

A higher dust-to-gas ratio leads to overall higher opacities because the opacities are only determined by the contribution of solids, resulting in 
higher disc temperatures and a higher aspect ratio. Likewise, a lower dust-to-gas ratio causes the aspect ratio to be smaller.

\subsection{Low viscosity discs}

A lower level of turbulence (i.e. a smaller $\alpha$ value) causes the disc to be colder. This is because of two effects: On one hand, a lower $\alpha$ viscosity enables the dust grain to grow to larger sizes (Eq. \ref{eq: max grain size}) as there are fewer destructive collisions in the disc, causing cooling to be more efficient. On the other hand, viscous heating, which is dominant in the inner part of the disc \citep{dullemond2001, 
bitsch2013}, becomes less efficient.

When utilising $\alpha$ viscosities of $10^{-3}$ and $5\times10^{-4}$ (Fig. \ref{fig: low viscosity}), the aspect ratio of the discs where the grains feature a constant dust fragmentation velocity of 1 m/s and that of the discs featuring a transition in the dust fragmentation velocity are qualitatively comparable to those of their counterparts with a high viscosity ($\alpha=5\times10^{-3}$; Fig. \ref{fig: high viscosity}). They show similar features in their radial profile, including those at the ice line, namely the bump in the disc with a constant dust fragmentation velocity and the sharp decrease in the disc featuring a transition in the dust fragmentation velocity.

In contrast to the case of $\alpha=5\times10^{-3}$, the aspect ratios of 
the discs with a dust fragmentation velocity of 5 m/s and with a transition from 1 m/s to 5 m/s no longer match for lower viscosities in the icy regions. This is caused by the fact that viscous heating is less efficient such that heating through stellar irradiation starts to dominate at smaller orbital distances, which can be seen by the radially increasing aspect ratio for the simulations that feature a constant dust fragmentation velocity of 5 
m/s \footnote{Stellar irradiation naturally gives a radially increasing aspect ratio \citep{chiang1997}.}. Due to the lower dust fragmentation velocity in the inner region of the discs that feature a transition in the dust 
fragmentation velocity, the discs are hotter, generating a bump in the aspect ratio. This casts a shadow onto the outer discs and results in lower temperatures in those outer discs, where a transition in the dust fragmentation velocity is employed, compared to the discs with a constant dust fragmentation velocity of 
5 m/s, where no inner bump is present.

 The profiles of the gas surface densities do not depend significantly on the value of the $\alpha$ viscosity, except for the position of the bump. 
This position changes with the location of the ice line in the underlying disc structure and is thus positioned closer to the star for a lower viscosity.

\subsection{Radial pressure gradients}

\begin{figure*}
    \centering
    \begin{subfigure}{0.33\textwidth}
        \includegraphics[width=\textwidth]{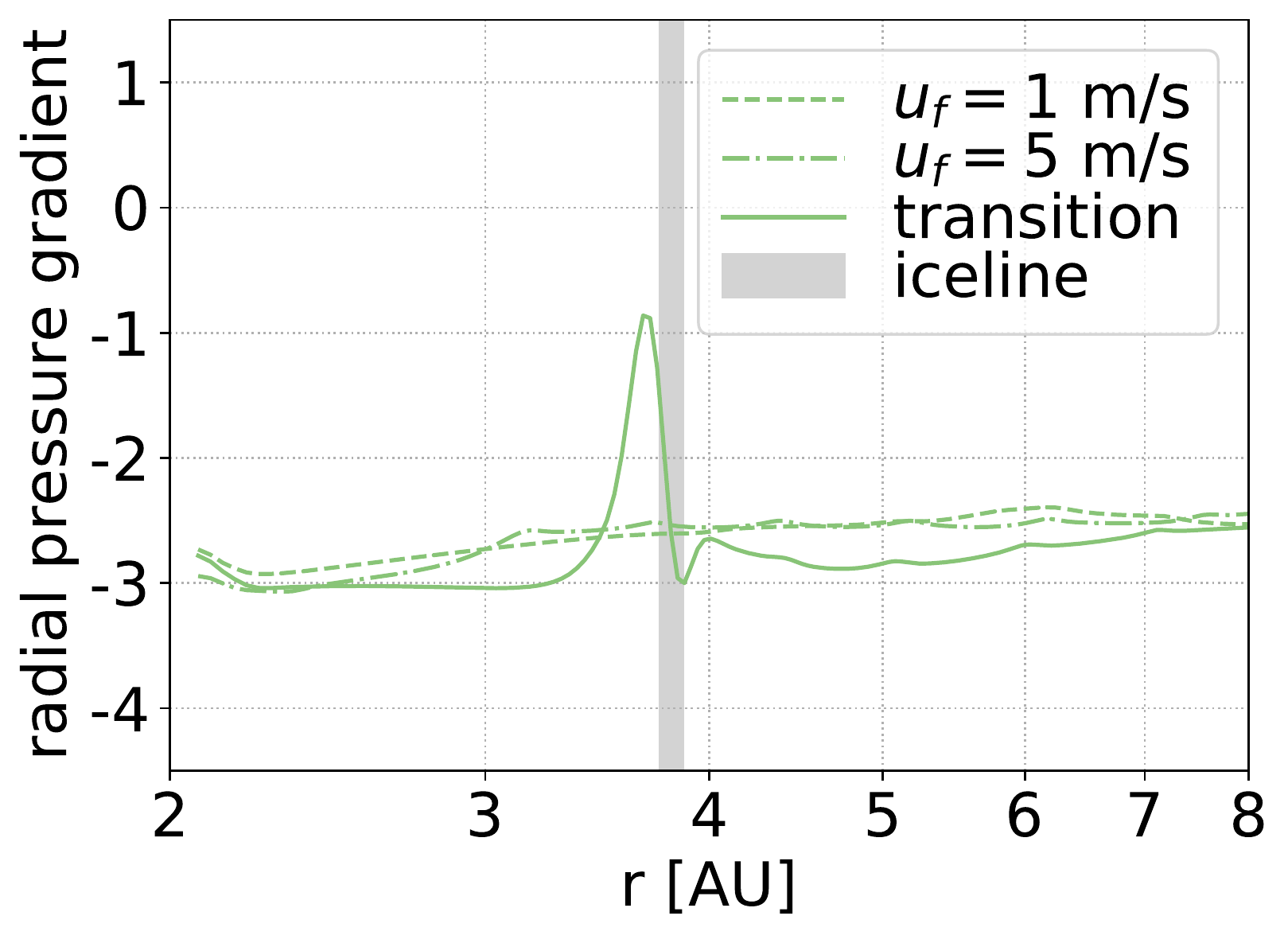}
        \caption{}
        \label{subfig: pressure gradient 5e-3}
    \end{subfigure}
    \begin{subfigure}{0.33\textwidth}
        \includegraphics[width=\textwidth]{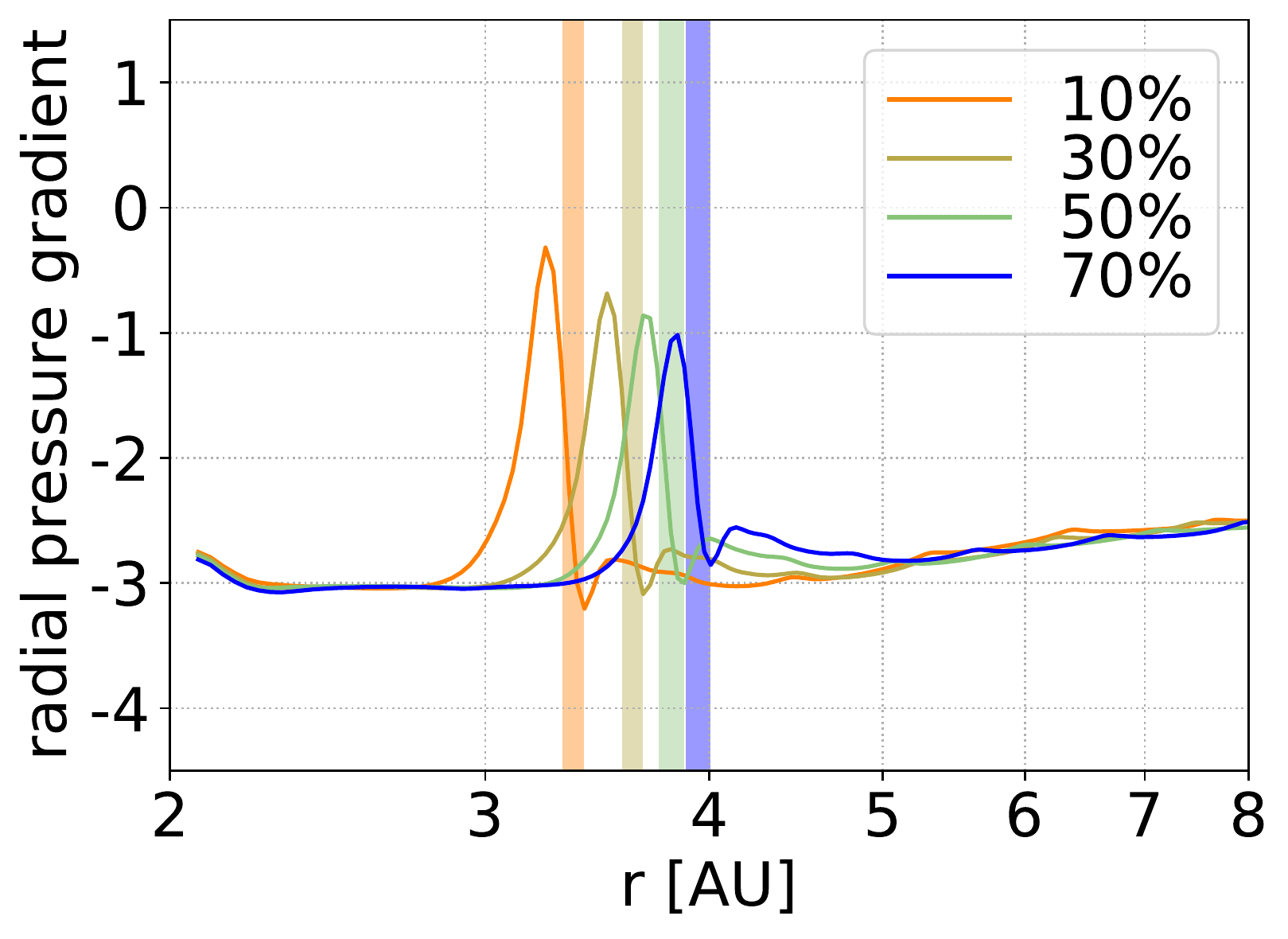}
        \caption{}
        \label{subfig: pressure gradient water content}
    \end{subfigure}
    \begin{subfigure}{0.33\textwidth}
        \includegraphics[width=\textwidth]{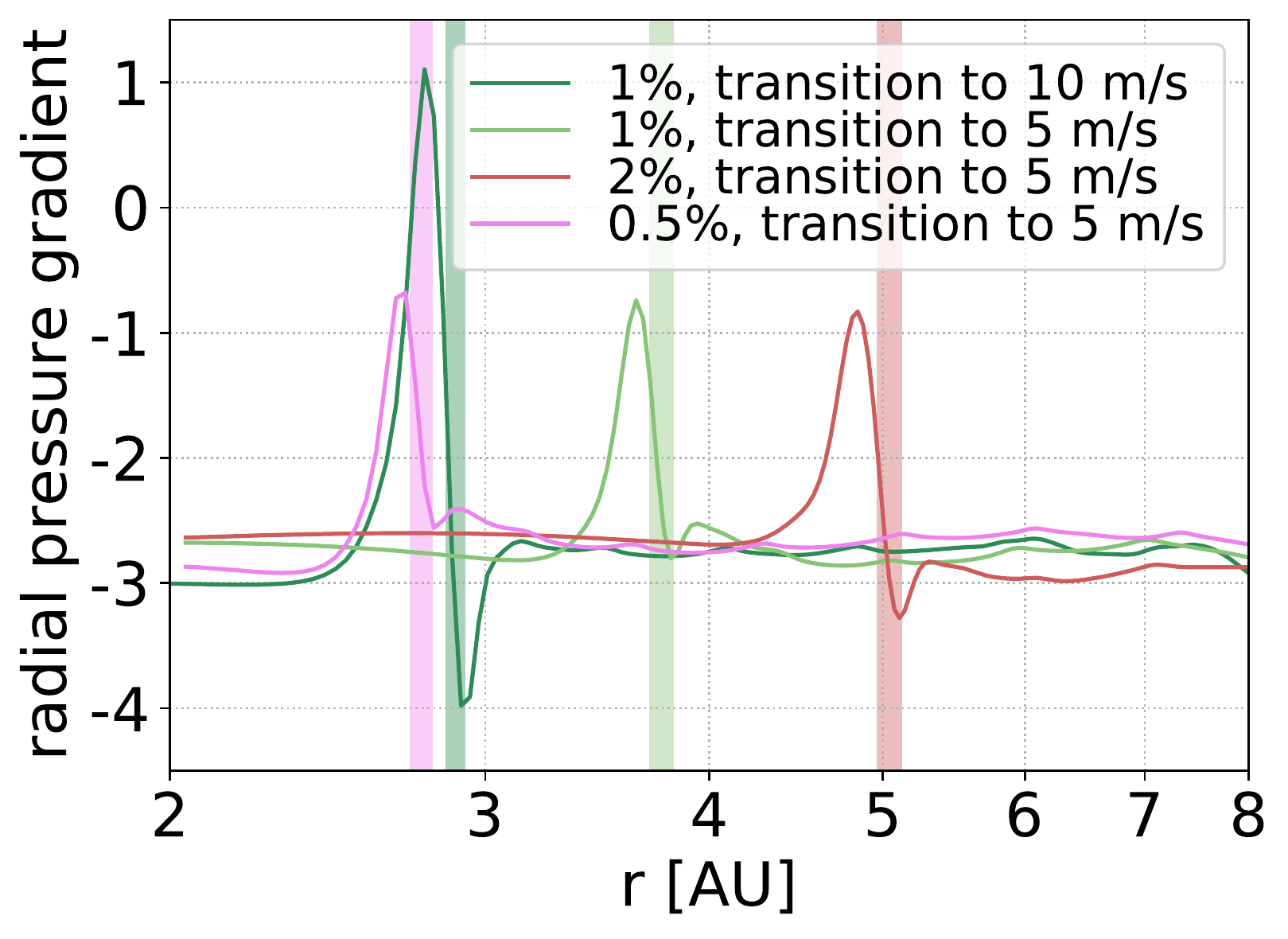}
        \caption{}
        \label{subfig: pressure gradient frag vel}
    \end{subfigure}
    \caption{Radial pressure gradient as a function of the orbital distance for discs with a viscosity of $\alpha=5\times10^{-3}$. In the left panel we show the radial pressure gradients that originate from the discs displayed in Fig. \ref{fig: high viscosity}. In the middle panel we show the pressure gradient of discs that feature a transition in the dust fragmentation velocity for different water abundances (Fig. \ref{subfig: water to silicate ratio AR}). In the right panel we show the pressure gradient of discs with a transition in the dust fragmentation velocity to 5 m/s and 
different dust-to-gas ratios, as well as a disc featuring a transition in the dust fragmentation velocity from 1 m/s to 10 m/s and a dust-to-gas ratio of 1\% (Fig. \ref{subfig: fragmentation velocity AR}). The coloured bars in Figs. \ref{subfig: pressure gradient water content} and \ref{subfig: pressure gradient frag vel} indicate the position of the ice line and are displayed in the colour of their respective disc parameters.}
    \label{fig: pressure gradient high viscosity}
\end{figure*}

\begin{figure*}
    \centering
    \begin{subfigure}{0.42\textwidth}
        \includegraphics[width=\textwidth]{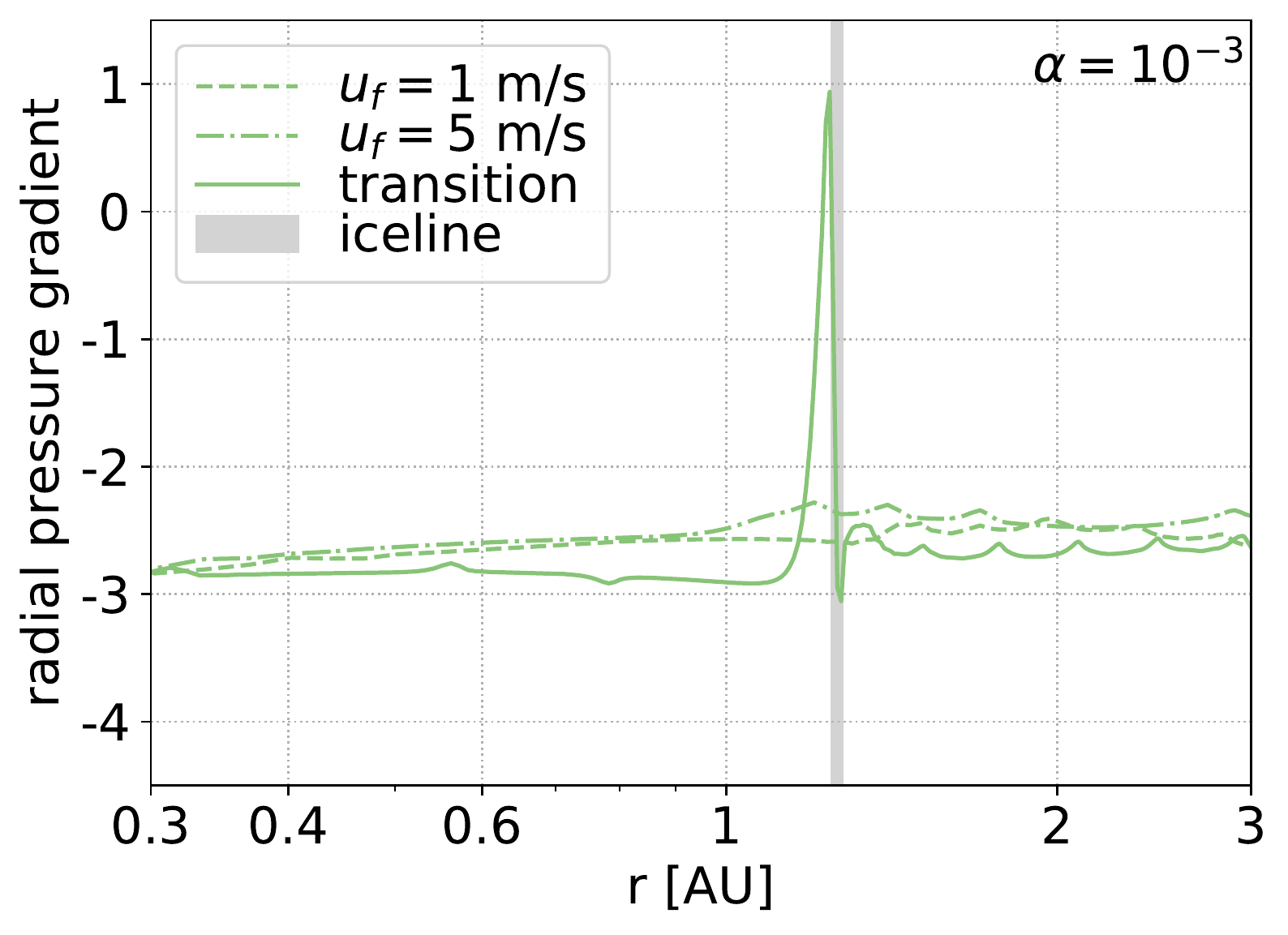}
    \end{subfigure}
    \begin{subfigure}{0.42\textwidth}
        \includegraphics[width=\textwidth]{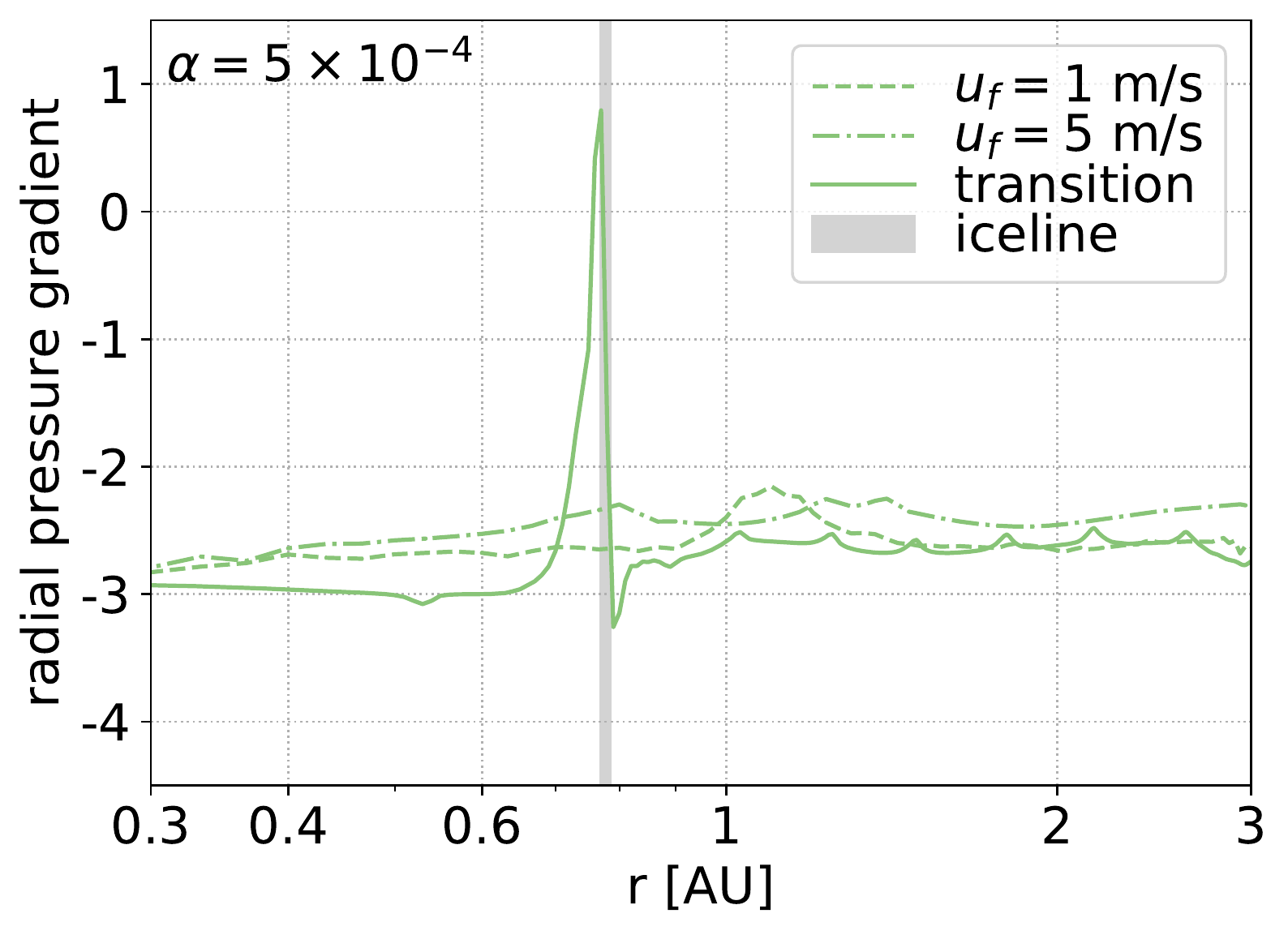}
    \end{subfigure}
    \caption{Radial pressure gradient as a function of the orbital distance with viscosity values of $\alpha=10^{-3}$ (left) and $\alpha=5\times10^{-4}$ (right). The pressure gradients shown here originate from the discs displayed in Fig. \ref{fig: low viscosity}.}
    \label{fig: pressure gradient low viscosity}
\end{figure*}

The radial pressure gradients of the discs with a high viscosity ($\alpha=5\times10^{-3}$) are displayed in Fig. \ref{fig: pressure gradient high viscosity}. We can see in Fig. \ref{subfig: pressure gradient 5e-3} that discs with a constant dust fragmentation velocity profile have a flat 
radial pressure gradient. The pressure gradient profile of the disc that features a transition in the dust fragmentation velocity, however, features a bump. The peak of this bump is located slightly inside the ice line. This is caused by the changes in the gas surface density and temperature profile at this location, which ultimately change the disc's pressure. It is present for all probed water abundances, but the orbital distance at which it is located changes with the water content in the disc (Fig. \ref{subfig: pressure gradient water content}). This can be traced back to the dependence of the position of the ice line on the water abundance of the underlying disc due to the different opacities of water-ice and silicate particles. In addition, the peak of the bump is higher for lower water abundances (meaning that the pressure gradient is locally smaller) because the transition at the ice line is sharper if the disc harbours a smaller amount 
of water.

A similar effect can be seen for the disc that features a transition in the dust fragmentation velocity from 1 m/s to 10 m/s instead of 5 m/s because, in this case, 
the ice line is located farther in  (Fig. \ref{subfig: pressure 
gradient frag vel}). Furthermore, the bump in the pressure gradient becomes more pronounced and even reaches positive values at the peak. This indicates the presence of a pressure bump in the disc and affects the radial 
distribution of the solids. We will discuss this matter and the implications of the pressure bump on planetesimal formation in Sect. \ref{discussion}. 

In Fig. \ref{subfig: pressure gradient frag vel} we also display the radial pressure gradients of the discs with different dust-to-gas ratios. Both the height and the shape of the peak are similar for all tested values. 
This is because the opacity inside and outside the ice line changes in the same way when varying the overall dust-to-gas ratio of the disc. The only relevant difference is the orbital distance at which the bump is located in its respective disc, which is directly connected to the location of the ice line.

In Fig. \ref{fig: pressure gradient low viscosity} we show the radial pressure gradients of the discs that feature a lower viscosity. We can see that 
the overall structure of the pressure gradients is similar to that of the 
discs with a high viscosity (Fig. \ref{subfig: pressure gradient 5e-3}), but two trends can be identified for a smaller $\alpha$ value: First, the pressure bump is located closer to the ice line, and secondly the peak of the bump is higher, even reaching positive values. This is because the transition region in the disc at the ice line covers a smaller radial range for a lower turbulence in the disc and is thus more pronounced (compare with Figs. \ref{fig: high viscosity} and \ref{fig: low viscosity}). We discuss the resolution dependence of our results in Appendix \ref{sec: resolution}.

In conclusion, we find that there is a bump in the radial pressure gradient profile for all tested parameters if the dust fragmentation velocity of the grains in the disc features a transition at the ice line. The bump is more pronounced for a lower viscosity, a lower water abundance, and a larger deviation in the dust fragmentation velocity inside and outside the ice line, while a change in the dust-to-gas ratio does not significantly affect the height of the bump.

\subsection{Planet migration}

\begin{figure}
    \resizebox{\hsize}{!}{\includegraphics{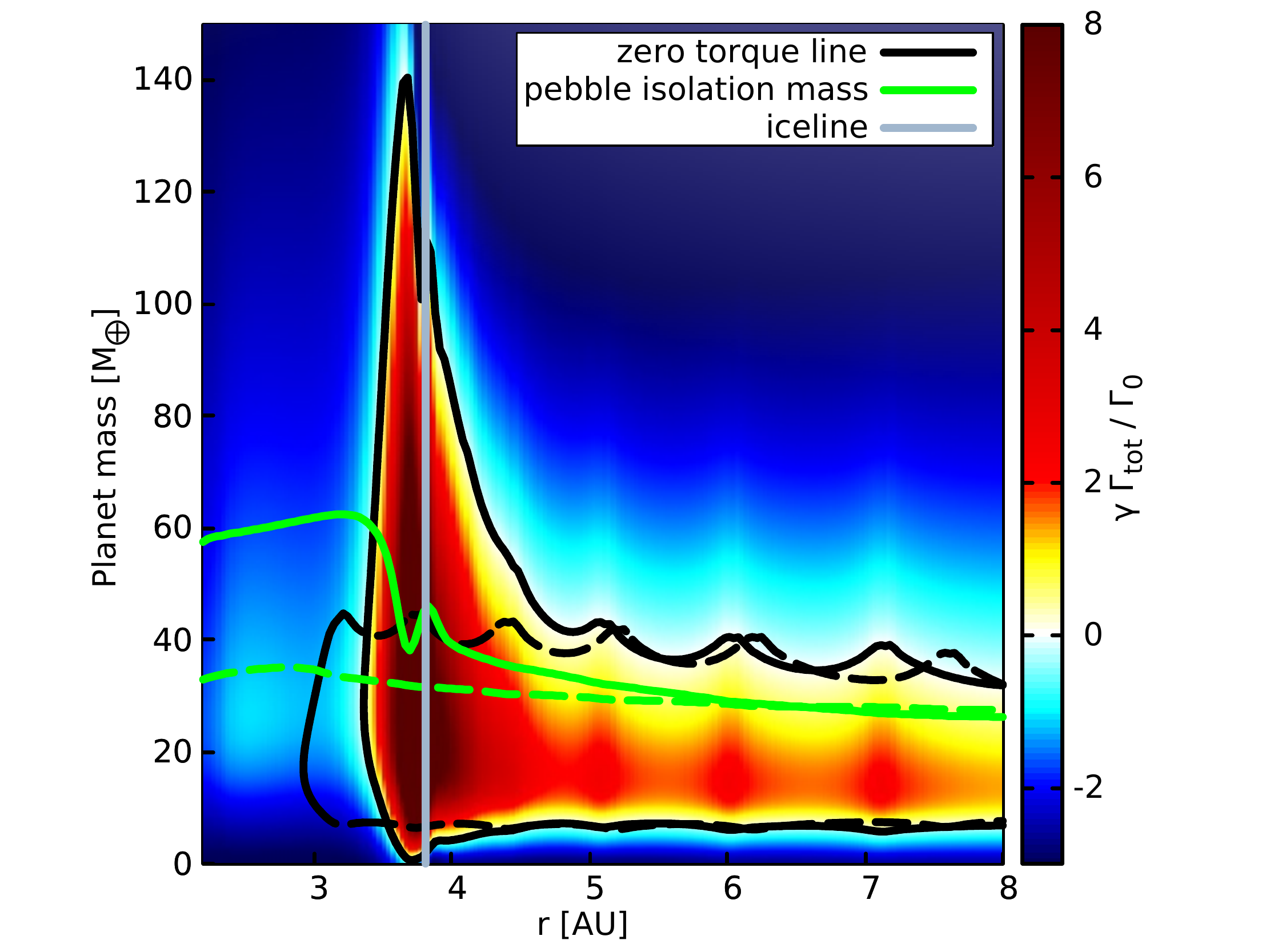}}
    \caption{Torque acting on planets in the disc that features a transition in the dust fragmentation velocity from 1 m/s to 5 m/s, $\alpha=5\times10^{-3}$, and a water abundance of 50\% (Figs. \ref{fig: high viscosity} and \ref{subfig: pressure gradient 5e-3}). The solid lines correspond to the disc that features the transition, and the dashed lines correspond to the disc that utilises the same disc parameters but has a constant dust fragmentation velocity of 5 m/s. The black lines are the zero torque lines, and the green 
lines represent the pebble isolation mass. The grey line marks the water-ice line of the disc in the case of a transition in the dust fragmentation 
velocity to 5 m/s.}
    \label{fig: mm 5e-3}
\end{figure}
\begin{figure}
    \resizebox{\hsize}{!}{\includegraphics{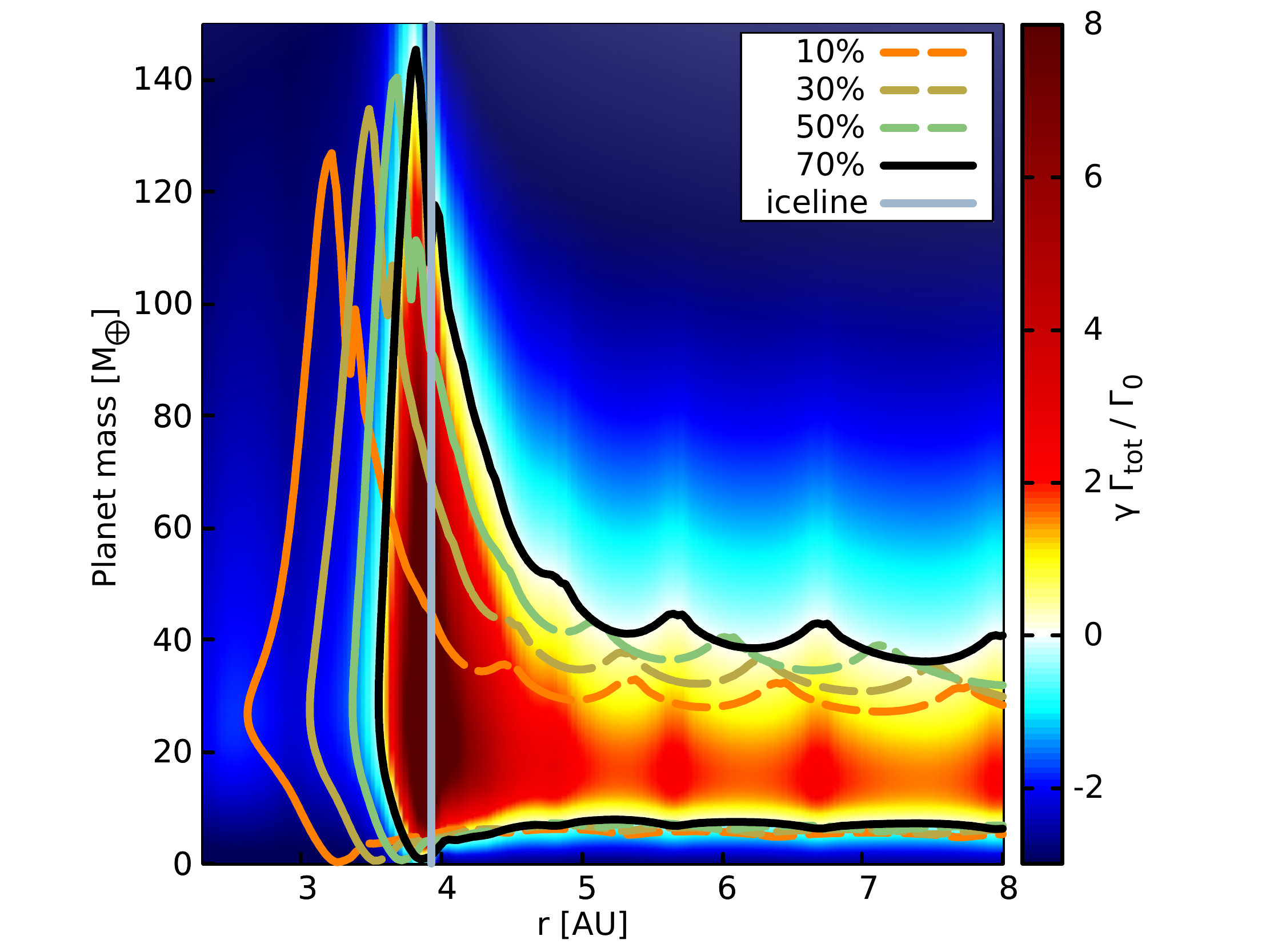}}
    \caption{Torque acting on planets in the disc that features a transition in the dust fragmentation velocity from 1 m/s to 5 m/s, $\alpha=5\times10^{-3}$, and a water abundance of 70\%. The black line is the zero torque 
line. The dashed coloured lines indicate the zero torque lines of discs with different water abundances, namely 10\%, 30\%, and 50\%, and $\alpha=5\times10^{-3}$ (Figs. \ref{subfig: water to silicate ratio AR} and \ref{subfig: pressure gradient water content}). All discs feature a dust-to-gas ratio of 1\%.}
    \label{fig: mm water content}
\end{figure}
\begin{figure}
    \resizebox{\hsize}{!}{\includegraphics{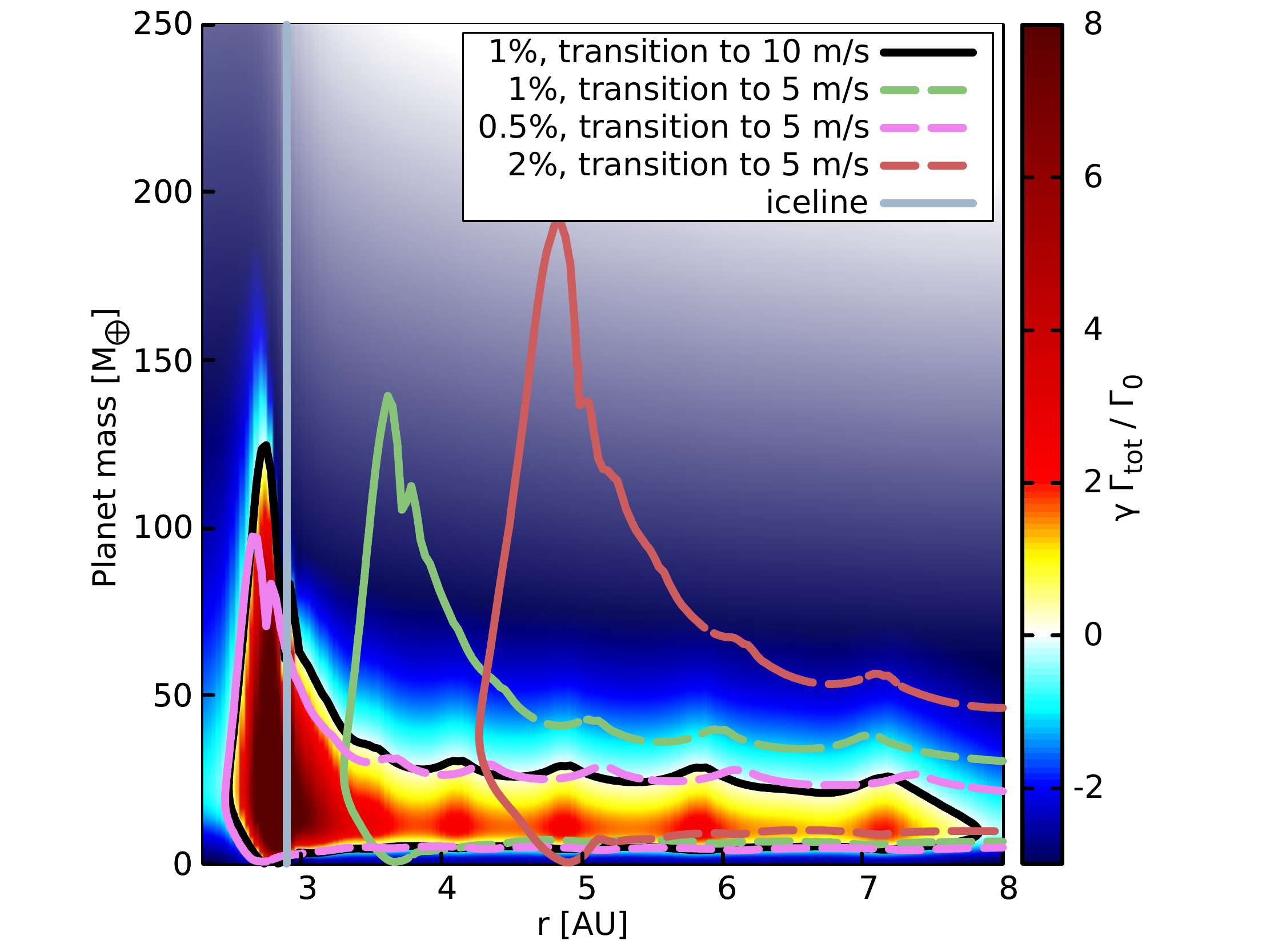}}
    \caption{Torque acting on planets in the disc that features a transition in the dust fragmentation velocity from 1 m/s to 10 m/s, $\alpha=5\times10^{-3}$, and a dust-to-gas ratio of 1\%. The black line is the zero torque line. The coloured lines indicate the zero torque lines of discs with a transition in the dust fragmentation velocity to 5 m/s for different dust-to-gas ratios, namely 0.5\%, 1\%, and 2\%, and $\alpha=5\times10^{-3}$ (Figs. \ref{subfig: fragmentation velocity AR} and \ref{subfig: pressure gradient frag vel}). All discs feature a water abundance of 50\%.}
    \label{fig: mm dust-to-gas}
\end{figure}

The thermal structure of the protoplanetary disc can influence the migration behaviour of the planets embedded in it (for a review, see \citealt{kley2012} and \citealt{baruteau2014}). Generally speaking, planets tend to migrate outwards when the local aspect ratio is a decreasing function of the orbital distance \citep{bitsch2013} if the viscosity is large enough to prevent saturation \citep{paardekooper2011}. Otherwise, they migrate inwards. Thus, planets can be trapped at the outer border of the zone of outward migration.

\begin{figure}
    \resizebox{\hsize}{!}{\includegraphics{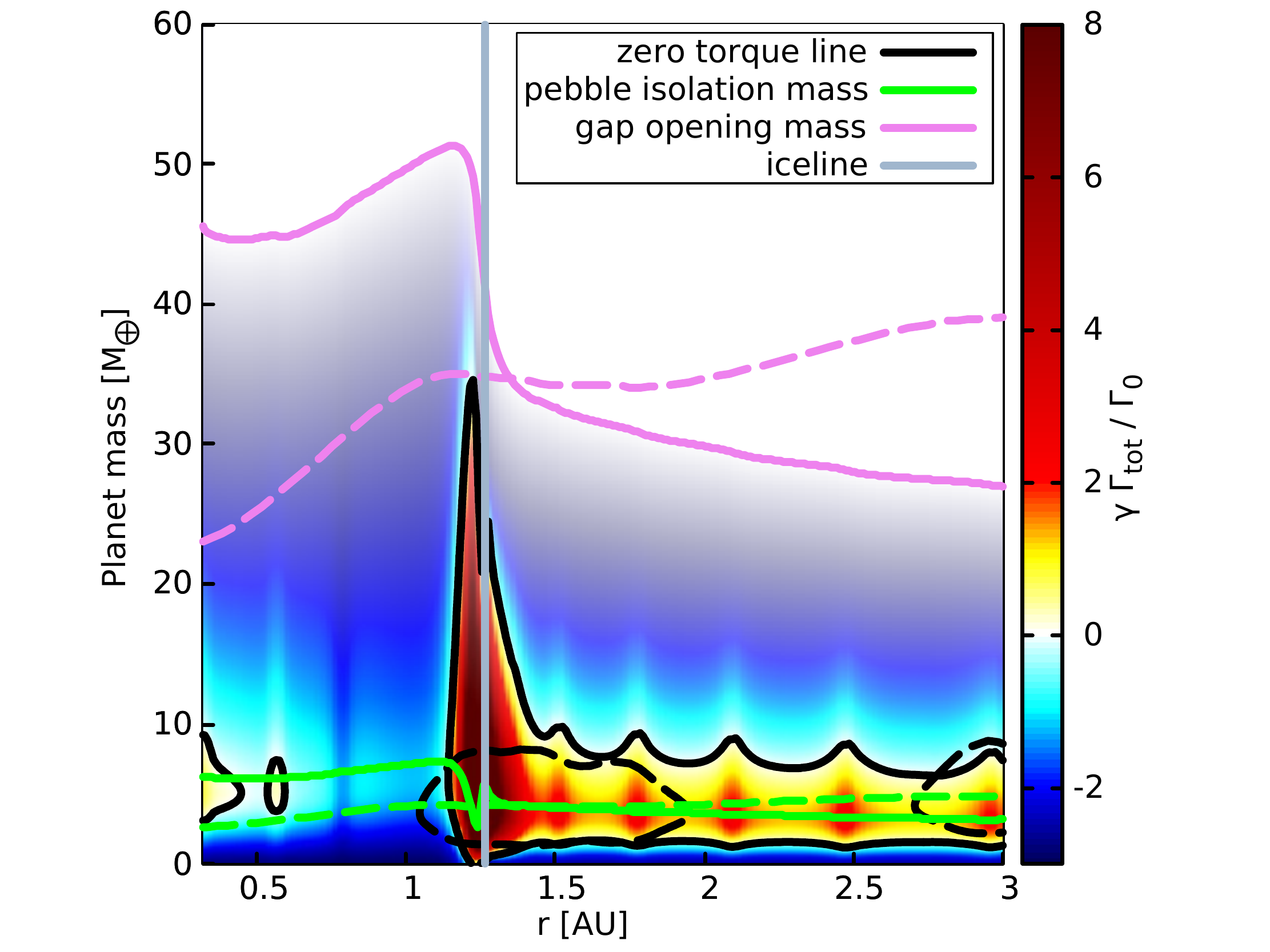}}
    \caption{Same as Fig. \ref{fig: mm 5e-3} but for discs with $\alpha=10^{-3}$ (Figs. \ref{fig: low viscosity} and \ref{fig: pressure gradient low 
viscosity}). We show in pink the gap opening mass, where the planets would transition to type-II migration.}
    \label{fig: mm 1e-3}
\end{figure}

\begin{figure}
    \resizebox{\hsize}{!}{\includegraphics{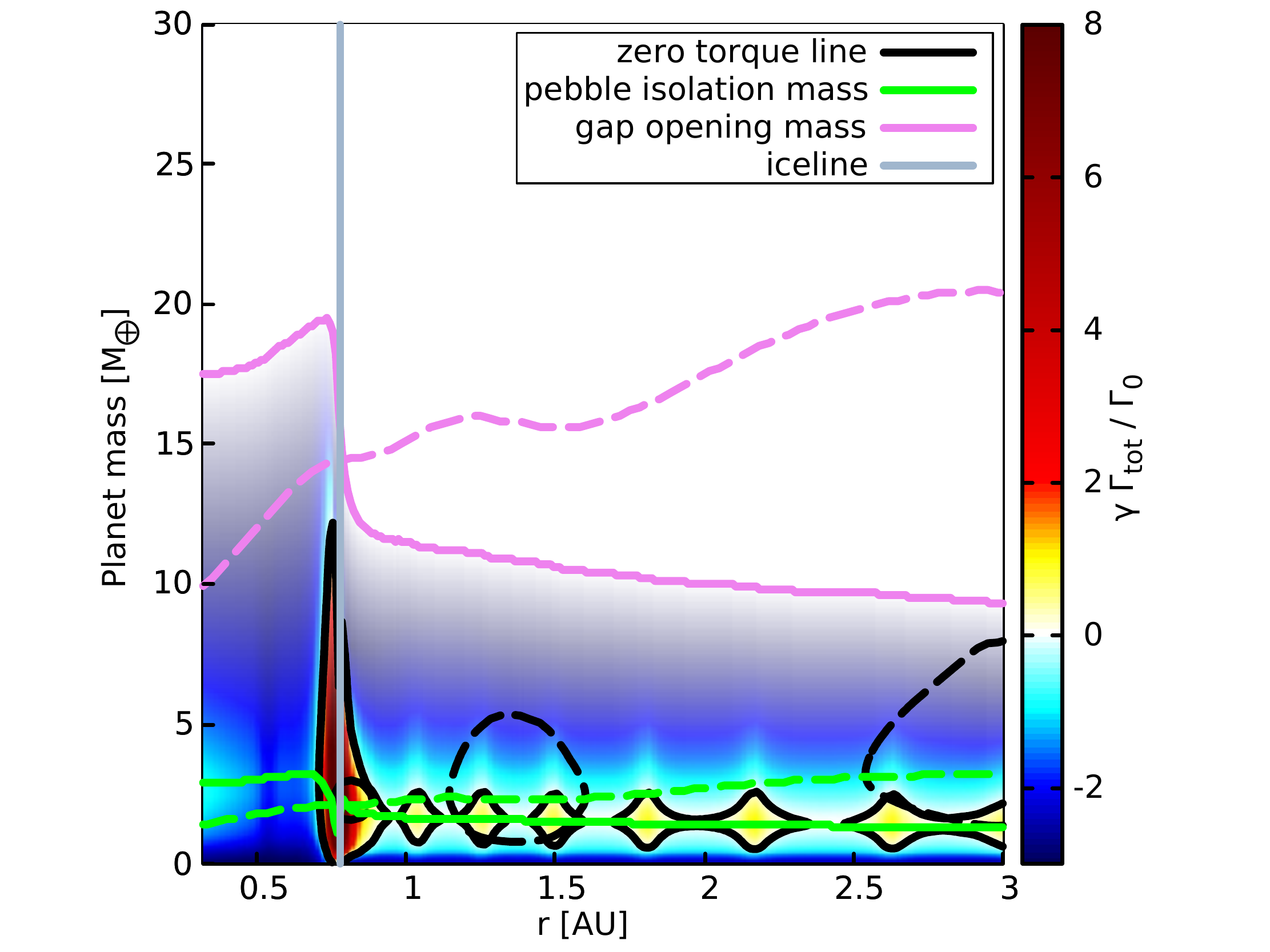}}
    \caption{Same as Fig. \ref{fig: mm 5e-3} but for discs with $\alpha=5\times10^{-4}$ (Figs. \ref{fig: low viscosity} and \ref{fig: pressure gradient low viscosity}). The pink line refers to the gap opening mass.}
    \label{fig: mm 5e-4}
\end{figure}

In order to portray the torque acting on the planets as a function of the 
orbital distance and the planetary mass, we present migration maps of some of the discs studied here. The migration rates are derived from the torque formula in \citet{paardekooper2011}. We show the migration maps for our studied discs in Figs. \ref{fig: mm 5e-3}, \ref{fig: mm water content}, \ref{fig: mm dust-to-gas}, \ref{fig: mm 1e-3}, and \ref{fig: mm 5e-4}. In these figures, we indicated the water-ice line at the 170 K mark with a vertical grey line. Furthermore, we include a line corresponding to the pebble isolation mass (at which the planet can start gas accretion) and another one referring to the gap opening mass (at which the planet opens a deep gap in the protoplanetary disc, indicating a transition to type-II migration).

For simplicity, we used the pebble isolation mass without diffusion from \citet{bitsch2018}, which is given by
\begin{gather}
\begin{split}
     M_{\text{iso}} = 25 \left[\frac{H/r}{0.05}\right]^3 \left[0.34\left(\frac{\log(10^{-3})}{\log(\alpha)}\right)^4 + 0.66\right]\\\
     \times \left[1 - \frac{\partial \ln P/\partial \ln r + 2.5}{6}\right]M_\oplus,
\end{split}
\end{gather}
where $M_\oplus$ denotes the mass of the Earth, $H/r$ the aspect ratio, and $\partial\ln P/\partial\ln r$ the radial pressure gradient.

The gap opening mass is determined by the following criterion, which indicates that a planet embedded in the disc would open a gap with a depth $\left(\text{i.e.}\ \frac{\Sigma_{G,\text{with planet}}}{\Sigma_{G,\text{without planet}}}\right)$ of maximum 10\% \citep{crida2006}:
\begin{gather}
    \frac{3}{4}H/r\left(\frac{3M_*}{M_P}\right)^{1/3} + 50\alpha (H/r)^2\frac{M_*}{M_P} \leq 1.
    \label{eq: gapmass}
\end{gather}
Here, $M_*$ is the stellar mass and $M_P$ is the planetary mass. We have assumed that the planet will transition smoothly from type-I to type-II migration between a gap depth of 50\% \citep{crida2007} and 10\%. We use a fading gradient of the colour scheme in the migration maps to indicate that the torque formula from \citet{paardekooper2011} might no longer be applicable once the planets start to open a deeper gap (50\% of the unperturbed 
gas surface density profile).

At first, we focused on the discs with a high viscosity ($\alpha=5\times10^{-3}$). In Fig. \ref{fig: mm 5e-3} we show the zero torque lines of the discs that feature a transition in the dust fragmentation velocity and 
a constant dust fragmentation velocity of 5 m/s. Beyond $\sim5$ AU, the regions of outward migration are similar, but near the ice line the disc with a transition in the dust fragmentation velocity features a band of strong outward migration that is not present in the disc with a constant dust 
fragmentation velocity. Therefore, planets with both lower and much higher masses than those affected by outward migration in the disc where grains have a constant dust fragmentation velocity can be trapped in 
the disc that features a transition in the dust fragmentation velocity. Additionally, growing planets reach the pebble isolation mass before they start migrating inwards, indicating that planets formed outside the 
water-ice line finish their assembly in these regions and become water-rich planets \citep{bitsch2019}.

For an increase in water abundance, the band of outward migration is located closer to the star because the disc is colder and the ice line is located farther in (Fig. \ref{fig: mm water content}). If the water content in the disc is smaller, the zone of outward migration only extends to 
smaller planetary masses. This is due to the aspect ratio of the underlying disc being smaller, which causes the torque to saturate for smaller planetary masses \citep{paardekooper2011}.

Planets embedded in the disc with a transition in the dust fragmentation velocity from 1 m/s to 10 m/s (Fig. \ref{fig: mm dust-to-gas}) show a similar migration behaviour, but the regions outside the ice line of the disc are colder compared to the disc that features the transition to 5 m/s.\ This only leads to outward migration for planets with lower masses.

The migration pattern of embedded low mass planets in a disc with a dust-to-gas ratio of 0.5\% is comparable to the migration pattern in a disc featuring a transition in the dust fragmentation velocity to 10 m/s outside the ice line. Inside the ice line, the smaller aspect ratio of the disc with $f_{DG}=0.5\%$ causes outward migration to saturate for lower mass planets compared to the disc with a transition to 10 m/s. In the disc featuring a dust-to-gas ratio of 2\%, on the other hand, the band of outward migration in located farther away from the star, and outward migration does not saturate until the planet reaches a very high mass. This is caused by the hot underlying disc, which leads to a high aspect ratio and the ice line being located at larger orbital distances.


The migration maps of the discs with a lower viscosity are displayed in Figs. \ref{fig: mm 1e-3} and \ref{fig: mm 5e-4}. They show the same trends as the discs with a higher viscosity (Fig. \ref{fig: mm 5e-3}), but since the aspect ratio of the discs is smaller, outward migration only affects lower mass planets. Another major difference is that the zero torque lines of the disc with a constant dust fragmentation velocity and the 
disc featuring a transition in the dust fragmentation velocity do not match outside the ice line (contrary to the high viscosity case), which is 
due to the deviation in the aspect ratio of the discs in the icy regions. 
As the aspect ratio of the low viscosity discs is smaller, the co-rotation 
torque saturates for lower planetary masses. In addition, the zone of outward migration is split up into two regions in the case of a constant dust fragmentation velocity profile. Due to the low viscosities and low aspect ratios, the planets transition to type-II migration at lower masses (Eq. \ref{eq: gapmass}). We mark this transition with the pink line in Figs. \ref{fig: mm 1e-3} and \ref{fig: mm 5e-4}.

We find an efficient migration trap located at the ice line for all tested 
parameters if the dust fragmentation velocity profile of the disc features a transition. The migration trap is able to catch planets with both lower and higher masses than those caught by the trap in a disc with a constant dust fragmentation velocity. It fixes the orbits of planets throughout a large span 
of masses, giving them time to reach the runaway gas accretion phase where $M_{\text{core}}\approx M_{\text{envelope}}$ before they migrate inwards. In particular, the planet trap located at the water-ice line is also present at low viscosities, where the entropy-driven co-rotation torque can usually no longer support outward migration in smooth discs. 

%%%%%

\section{Discussion} \label{discussion}

\subsection{Pressure trap}

If the solids in the protoplanetary disc have different dust fragmentation velocities inside and outside the ice line, the radial pressure gradient features a peak in its profile and locally reaches smaller values than in discs that have a constant dust fragmentation velocity profile. For a large variance in the dust fragmentation velocity or a small $\alpha$-viscosity value, the radial pressure gradient can even hold positive values, indicating the presence of a pressure bump. Since radial grain drift \citep{weidenschilling1977, brauer2008} converges from both the inner and outer side of the disc towards the pressure maximum \citep{johansen2007nature}, the solids accumulate in the pressure bump, causing a pileup. However, radial grain drift was not considered in our simulations for the sake of simplicity, and thus the pileup is not simulated. If radial grain drift were included, we would expect the pileup of solids to locally increase the dust density, resulting in an increase in the opacity inside the water-ice line. This would enhance heating and diminish cooling processes in the disc, leading to a higher local disc temperature and a larger aspect ratio 
inside the water-ice line. Based on this, we anticipate that the transition at the ice line would be more pronounced if the radial grain drift that leads to pileups inside the water-ice line \citep{drazkowska2016} were considered.

\subsection{Planetesimal formation}

Planetesimals are thought to be born big, typically with a size of 100 km 
\citep{bottke2005, morbidelli2009}. They are expected to form through collapses of pebble clumps that are caused by processes such as the streaming instability \citep{youdin2005, johansen2007, johansen2007nature, klahr2020}. This process requires an elevated solid-to-gas ratio, where the exact solid-to-gas ratio needed to form planetesimals depends on the particle size \citep{carrera2015, yang2017}. \citet{bai2010} found that the streaming instability is more efficient when the aspect ratio of the disc and 
the local radial pressure gradient are smaller. In a disc that features a 
transition in the dust fragmentation velocity at the ice line, the aspect ratio decreases rapidly in the transition region and the profile of the radial pressure gradient features a bump that is located slightly inside the ice line. Our simulations show that the height of the bump is not affected by the dust-to-gas ratio of the disc and only varies in its radial position, corresponding to the location of the ice line in the disc. We assume that this is also true for evolved discs that might feature a gradient in the dust-to-gas ratio due to radial drift because the ice line covers a radially small region where the number of solids would only undergo a small change. Furthermore, the dust-to-gas ratio in these evolved discs would be higher inside the ice line than outside it, causing the transition at the ice line to be even more prominent. Therefore, the conditions we find at the location of the ice line are favourable for the streaming instability to be triggered, enhancing or even initiating planetesimal formation around the water-ice line.

\subsection{Migration trap}

The migration of planets in the protoplanetary disc is determined by the thermal structure and the viscosity of the disc. A transition in the dust fragmentation velocity at the ice line greatly influences the thermal structure and also affects the migration behaviour of embedded planets. In a disc that features such a transition in the dust fragmentation velocity, the 
zone of outward migration features a radially narrow band at the ice line, which extends from low to very high planetary masses and does not exist in discs that have a constant dust fragmentation velocity profile. It is present for all tested viscosity values and provides an efficient migration trap, enabling the planets to reach runaway gas accretion \citep{pollack1996} before migrating into the inner regions. If a transition in the dust fragmentation velocity is present at the ice line, it will have important consequences for planet formation as planets forming in the outer disc do not migrate inwards during their envelope contraction phase \citep{bitsch2015}, but rather at a later stage when they are in a runaway gas regime.

We have assumed that the presence of a planet with a low mass undergoing type-I migration does not disturb the disc’s overall structure enough to destroy the pressure and migration trap at the ice line. However, this might not be true for massive planets, which may even 
be able to change the temperature of the disc and reshape the water-ice line \citep{ziampras2020}. The influence of a massive planet on the pressure and migration trap at the ice line in discs that feature a transition in the dust fragmentation velocity is beyond the scope of the present work and will be investigated in a future study.

\subsection{Planetary composition}

Small super-Earth planets could have predominately rocky compositions \citep{fulton2017, owen2017, jin2018}, while larger mini-Neptunes with planetary envelopes could harbour a significant amount of water \citep{venturini2020}. Planetary embryos starting to grow outside the water-ice line 
will accrete water-rich material, forming water-rich planetary cores \citep{bitsch2019, izidoro2019, venturini2020}. The water-rich planets would have to migrate inwards close to the central star to make up the population of water-rich super-Earths \citep{bitsch2019, venturini2020}. This inward migration is normally possible in discs with low viscosities because the migration trap at the water-ice line is not efficient \citep{bitsch2019}. However, if the migration trap at the water-ice line is very efficient, as our study shows, single close-in water-rich super-Earths cannot just smoothly migrate inwards; however, they could potentially do so in a convoy of super-Earths or at later disc evolution stages \citep{izidoro2017, izidoro2019}.

%%%%%

\section{Summary} \label{summary}

In this paper we have studied how a transition in the fragmentation velocity of dust particles at the ice line impacts the structure of the protoplanetary disc, the formation of planetesimals, and the migration behaviour of planets. We tested different dust fragmentation velocity values, $\alpha$-viscosity values, dust-to-gas ratios, and water abundances. A larger dust 
fragmentation velocity and a lower $\alpha$ value enable the dust grains 
to grow to larger sizes, causing them to have lower opacities and leading 
to lower disc temperatures and a smaller aspect ratio. A change in the water content of the disc affects the opacities inside and outside
the ice line and changes the position of the ice line. For lower water abundances, the ice line is located closer to the star (Fig. \ref{subfig: water to silicate ratio AR}). An increase in the number of solids present in the disc leads to higher opacities and overall hotter discs, causing the ice line to be located farther away from the star. The opposite is true for a decrease in the dust-to-gas ratio (Fig. \ref{subfig: fragmentation velocity AR}).

In discs that feature a transition in the dust fragmentation velocity at the ice line, we found that the aspect ratio decreases rapidly in the region around the ice line, while the gas surface density increases. The gas surface density is rearranged due to changes in the viscosity caused by the changes in the aspect ratio and the temperature of the disc. This is true for all probed parameters. Additionally, the profile of the radial pressure gradient features a bump, which tends to be positioned slightly inside the ice line. Due to the locally smaller radial pressure gradient, planetesimal formation via the streaming instability is expected to be enhanced or even initiated at that location. For large deviations in the dust 
fragmentation velocity values of the grains in the hot and icy regions of 
the disc and in discs with a small level of turbulence, we identified a pressure bump near the ice line (Fig. \ref{subfig: pressure gradient frag vel}).

A transition in the dust fragmentation velocity also helps in trapping planets. In these discs, the zone of outward migration features a narrow band at the ice line, which covers a huge range of planetary masses, including 
masses lower than those affected by outward migration in the discs featuring a constant dust fragmentation velocity profile (e.g. Fig. \ref{fig: mm 5e-3}). Due to this band, the planets can reach runaway gas accretion before migrating into the hot regions of the disc. In particular, this planetary trap also exists at low disc viscosities, where outward migration in smooth discs is not possible.

We have shown that if the dust fragmentation velocity of water-ice and silicate grains is different, the ice line can serve as a pressure and migration trap. Therefore, we suggest that a constant, non-temperature-dependent dust fragmentation velocity profile is not sufficient to model the thermal structure of the protoplanetary disc, provided that the dust fragmentation velocity of water-ice and silicate grains is not equal.

Our models further strengthens the argument that the water-ice line could be a prime location for planetesimal and planet formation.

\begin{acknowledgements}
J.M., S.S., and B.B. thank the European Research Council (ERC Starting Grant 757448-PAMDORA) for their financial support. S.S is a Fellow of the International Max Planck Research School for Astronomy and Cosmic Physics at the University of Heidelberg (IMPRS-HD). In addition, we thank the anonymous referee for the valuable and constructive comments that helped to improve the manuscript.
\end{acknowledgements}

\bibliographystyle{aa}
\bibliography{ref}

\begin{appendix}

\section{Convergence and radial resolution}\label{sec: resolution}

\begin{figure*}[!ht]
    \centering
    \begin{subfigure}{0.42\textwidth}
        \includegraphics[width=\textwidth]{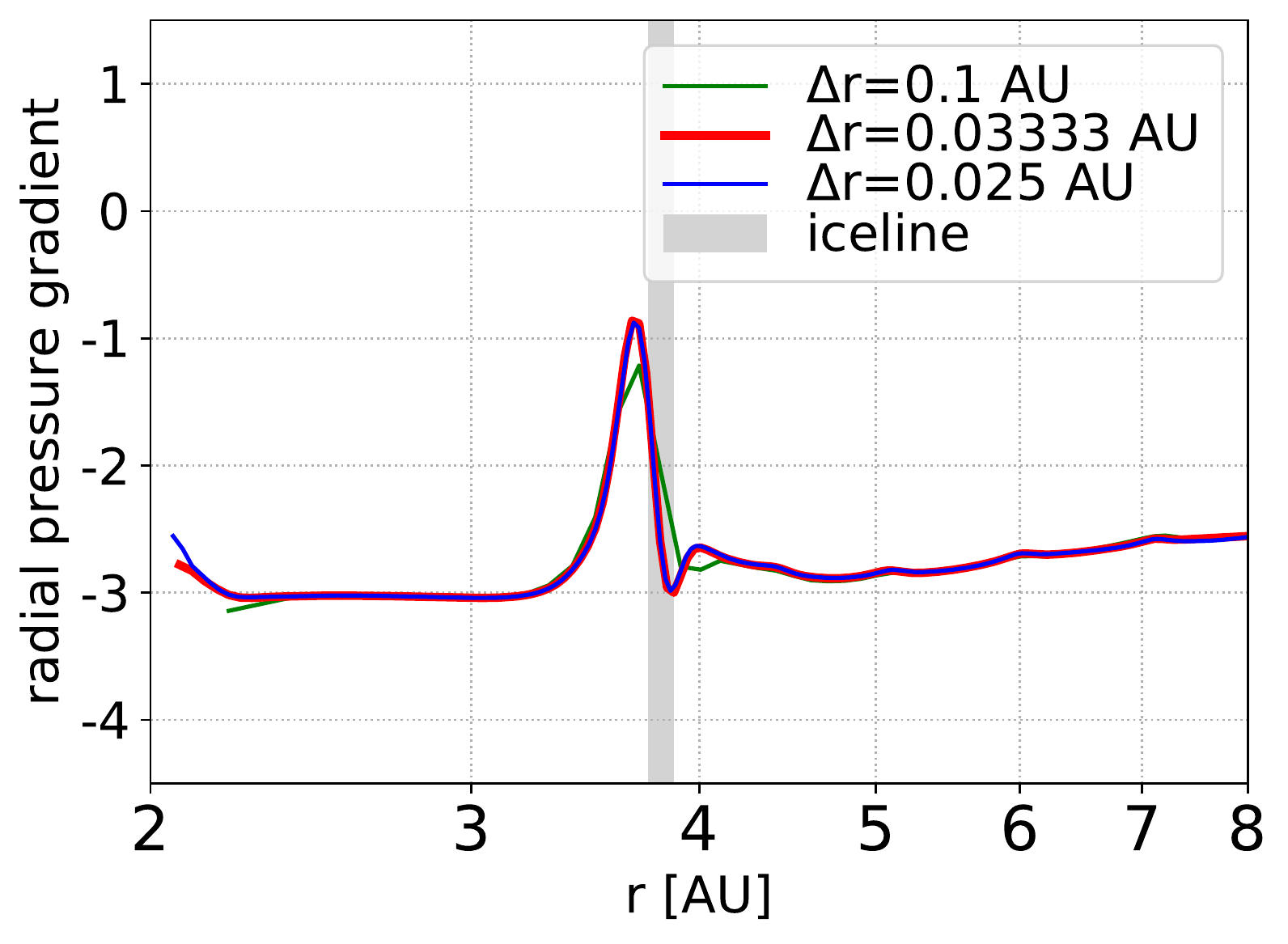}
        \caption{$\alpha=5\times 10^{-3}$, transition to 5 m/s}
    \end{subfigure}
    \begin{subfigure}{0.42\textwidth}
        \includegraphics[width=\textwidth]{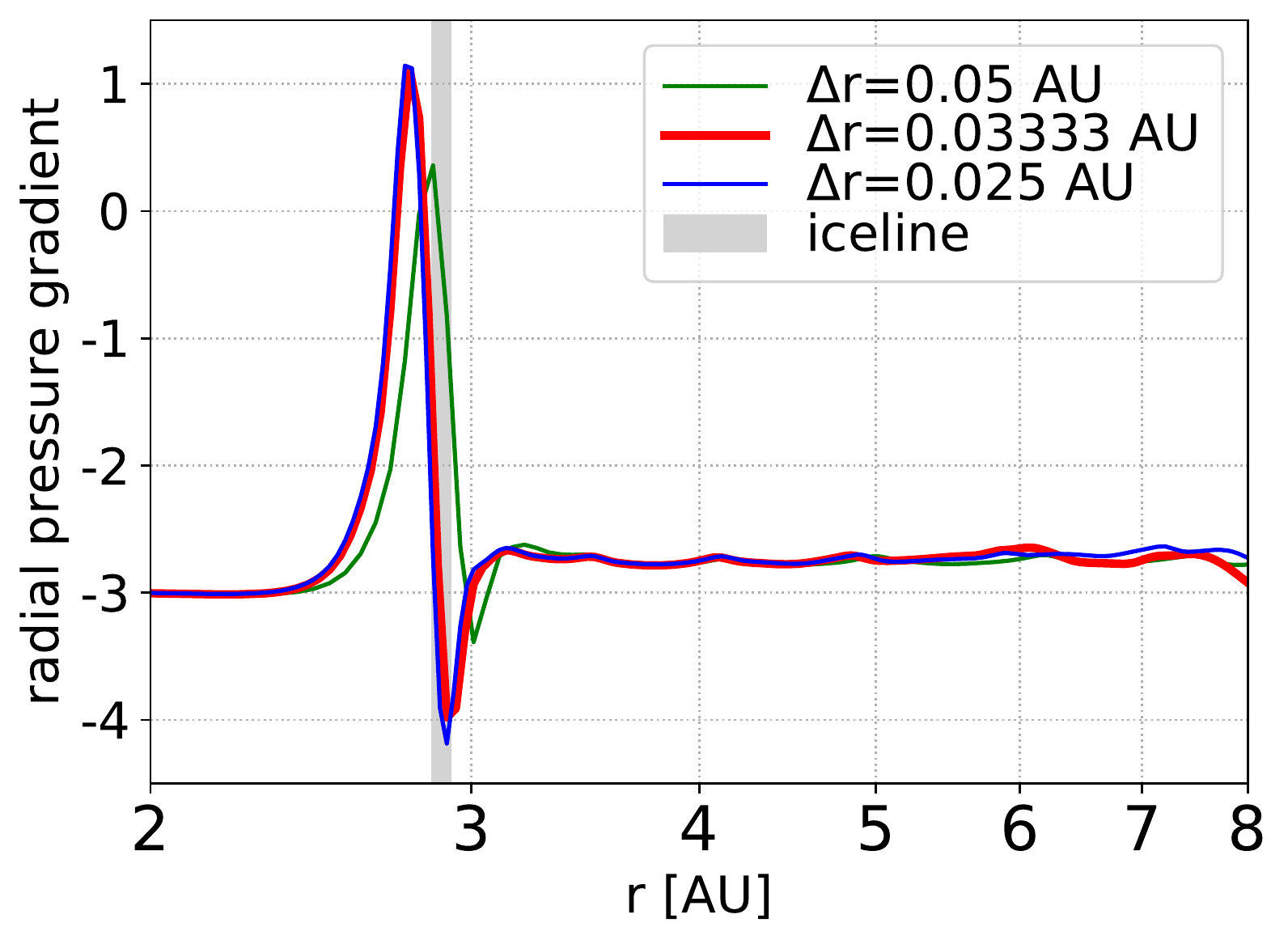}
        \caption{$\alpha=5\times 10^{-3}$, transition to 10 m/s}
    \end{subfigure}
    \begin{subfigure}{0.42\textwidth}
        \includegraphics[width=\textwidth]{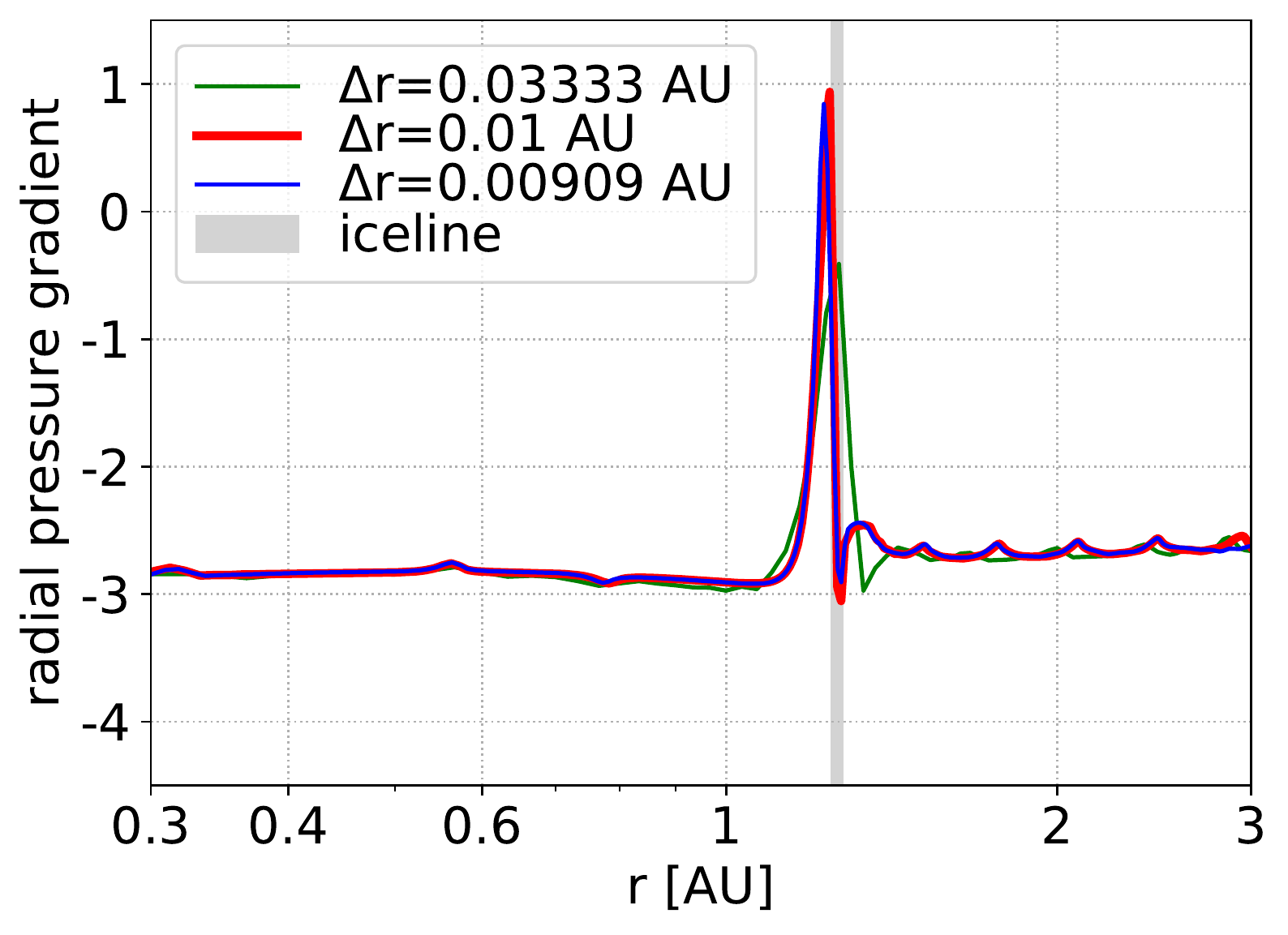}
        \caption{$\alpha=10^{-3}$, transition to 5 m/s}
    \end{subfigure}
    \begin{subfigure}{0.42\textwidth}
        \includegraphics[width=\textwidth]{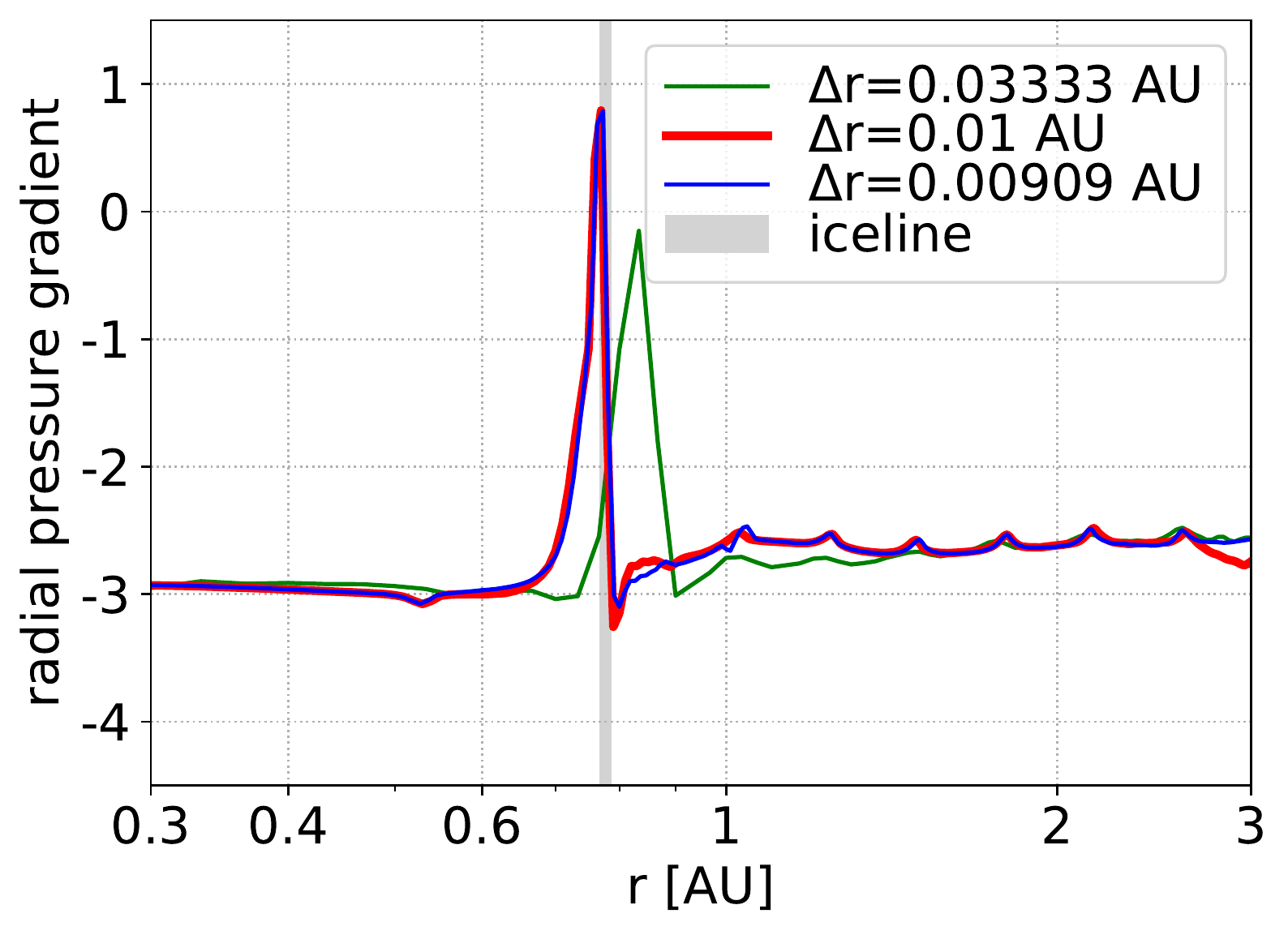}
        \caption{$\alpha=5\times 10^{-4}$, transition to 5 m/s}
    \end{subfigure}    
    \caption{Radial pressure gradient as a function of orbital distance for different radial resolutions. All discs feature a water abundance of 50\%, a dust-to-gas ratio of 1\%, and a transition in the dust fragmentation velocity. The resolutions that have been chosen in the present work are displayed in red.}
    \label{fig: convergence}
\end{figure*}

In this section we describe how the radial resolution of the simulations affects the structure of the discs that feature a transition in the dust fragmentation velocity.

The radial range in which the midplane temperature drops from 180 K to 160 K can be very narrow and thus demands a sufficiently high number of radial grid cells to be properly resolved. However, a higher radial resolution leads to much longer computation times. Therefore, it is necessary to find a compromise between radial resolution and time expenditure.

The radial resolution plays a crucial role in both the location of the ice line in the disc and the height of the peak of the radial pressure gradient, especially for the simulations that feature a lower viscosity. Because 
of this, we compared the radial pressure gradient of discs with different radial resolutions as a sensitive measure for convergence.

In Fig. \ref{fig: convergence} we show the radial pressure gradient of discs with a transition in the dust fragmentation velocity for different radial resolutions. All simulations feature a water abundance of 50\% and a dust-to-gas ratio of 1\%. The resolutions that have been chosen in the present work are displayed in red. All simulations converge for the chosen radial resolution.

Therefore, we anticipate that the physical structure of the discs would not qualitatively change for higher radial resolutions, even in the low viscosity case.

%%%%%

\section{Additional plots} \label{further plots}

Here we provide additional plots of the maximal grain size, the dust surface density, the vertical temperature profile, and the radial pressure profile of the disc that features $\alpha = 5\times 10^{-3}$, a water abundance of 50\%, a solid-to-gas ratio of 1\%, and a transition in the dust fragmentation velocity from 1 m/s to 5 m/s (Fig. \ref{fig: high viscosity}).

\begin{figure}
    \resizebox{\hsize}{!}{\includegraphics{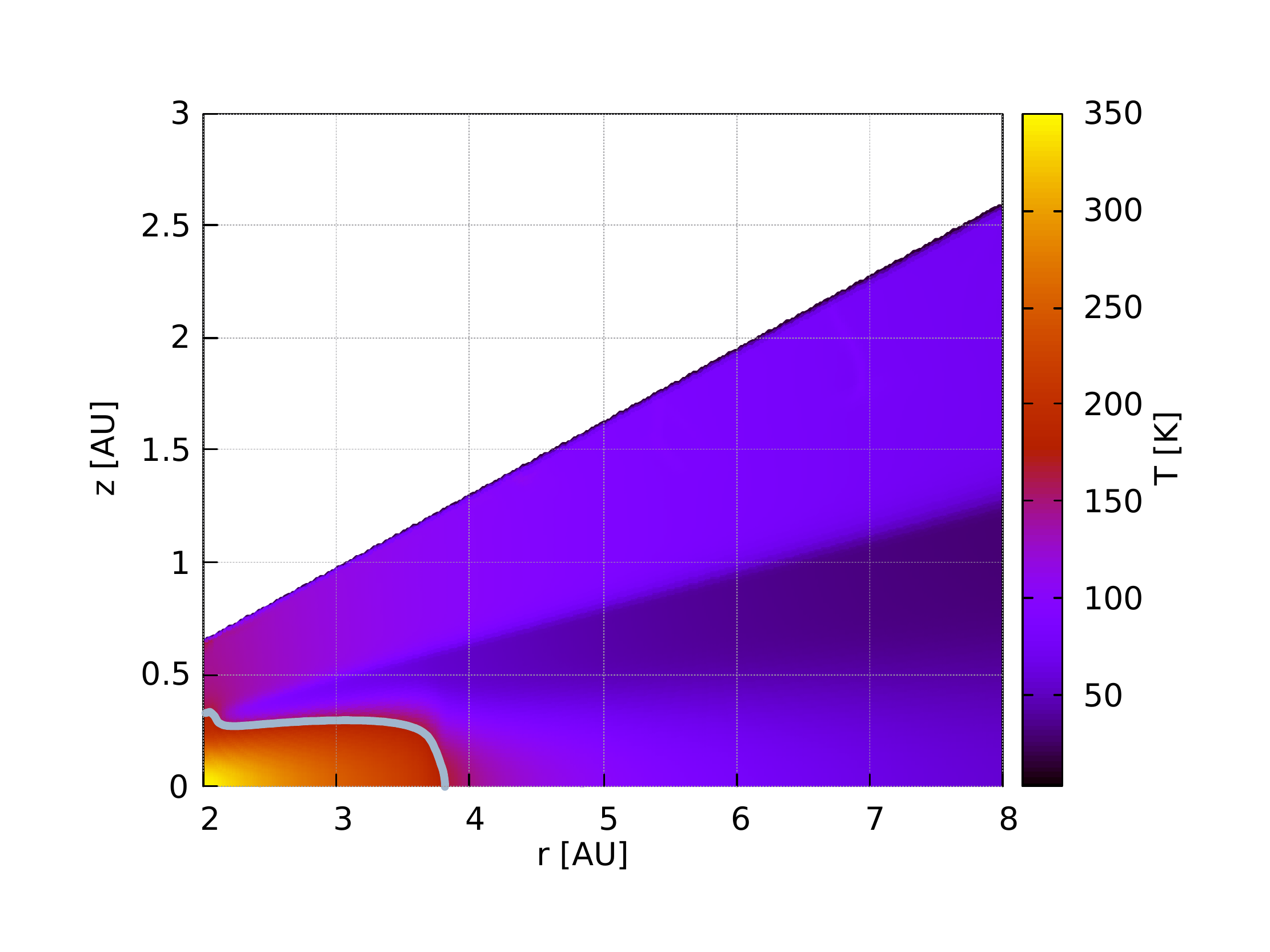}}
    \caption{Disc temperature as a function of orbital distance and vertical distance to the midplane for the disc that features $\alpha = 5\times 10^{-3}$, a water abundance of 50\%, a solid-to-gas ratio of 1\%, and a transition in the dust fragmentation velocity from 1 m/s to 5 m/s (Fig. \ref{fig: high viscosity}). The ice line is indicated as a grey line. The disc is vertically isothermal in the outer regions (not shown), where stellar irradiation dominates over viscous heating.}
\end{figure}
\begin{figure}
    \resizebox{\hsize}{!}{\includegraphics{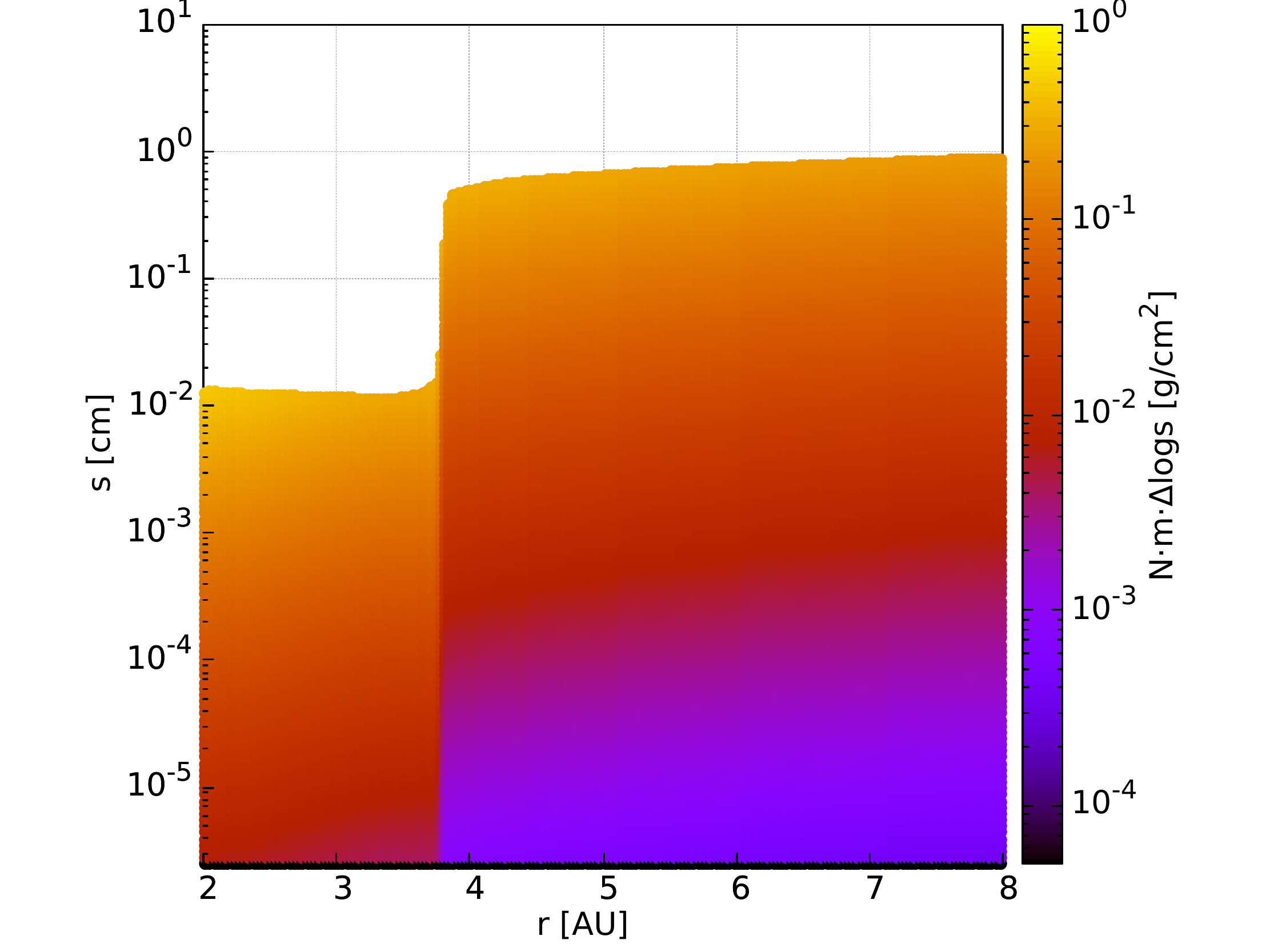}}
    \caption{Dust surface density as a function of orbital distance and grain size for the disc that features $\alpha = 5\times 10^{-3}$, a water abundance of 50\%, a solid-to-gas ratio of 1\%, and a transition in the dust fragmentation velocity from 1 m/s to 5 m/s (Fig. \ref{fig: high viscosity}). At the location of the ice line, the maximal grain size rapidly increases by a factor of $\sim40$. Since the maximum grain size is proportional to $u_f^2$ (Eq. \ref{eq: max grain size}), the transition in the dust fragmentation velocity from 1 m/s to 5 m/s corresponds to a change by a factor of 25. The remaining increase can be related to the gas surface density and temperature dependence of the maximum grain size. At the ice line there is a pronounced drop in temperature, which is responsible for the sharp decrease in the aspect ratio.}
\end{figure}
\begin{figure}
    \resizebox{0.9\hsize}{!}{\includegraphics{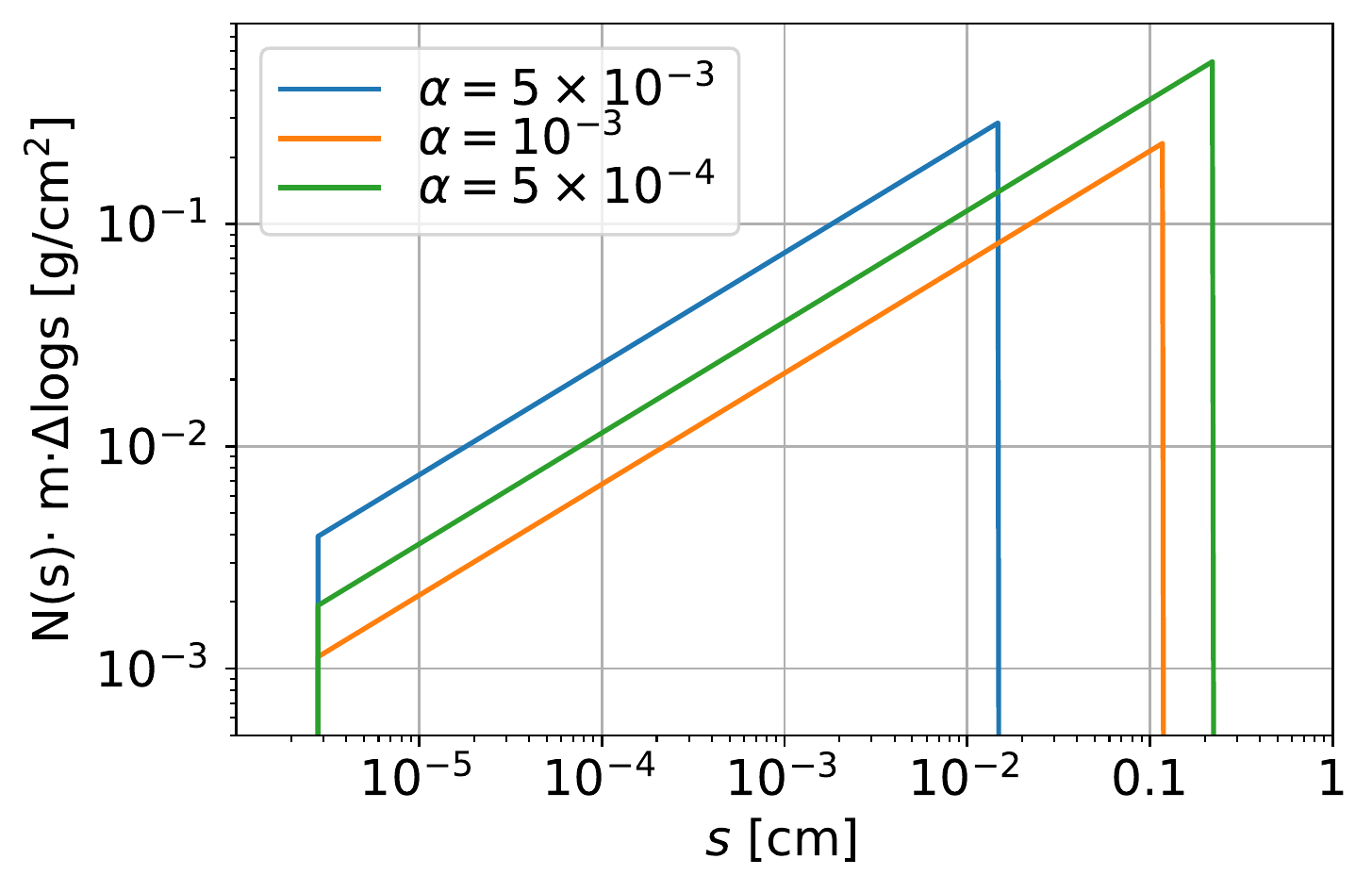}}
    \caption{Dust surface density as a function of the grain size for different turbulence parameters. All discs feature a water abundance of 50\%, a dust-to-gas ratio of 1\%, and a dust fragmentation velocity of 1 m/s. 
The dust surface density profiles are taken at a temperature of $\sim180$ 
K.
    The minimum grain size is 0.025 $\mu$m; the maximum grain size is computed using Eq. \ref{eq: max grain size}. The total dust surface density at each orbital distance is determined by the local gas surface density and the dust-to-gas ratio.}
\end{figure}
\begin{figure}
    \resizebox{0.9\hsize}{!}{\includegraphics{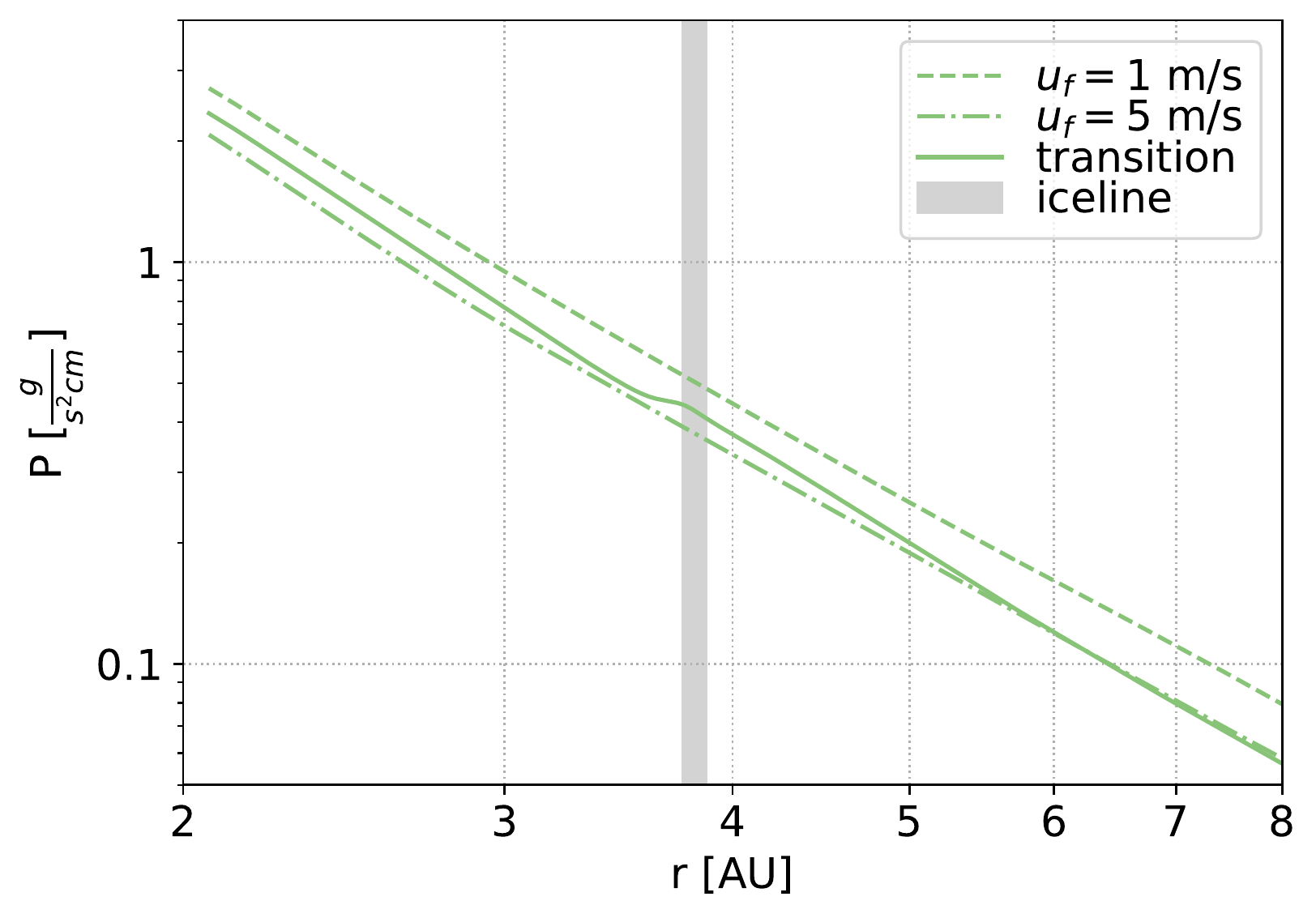}}
    \caption{Midplane pressure as a function of orbital distance for the disc that features $\alpha = 5\times 10^{-3}$, a water abundance of 50\%, a 
solid-to-gas ratio of 1\%, and a transition in the dust fragmentation velocity from 1 m/s to 5 m/s (Fig. \ref{fig: high viscosity}). The discs that feature constant dust fragmentation velocities of 1 m/s and 5 m/s are shown 
for comparison. The difference in pressure of the disc that features a constant dust fragmentation velocity of 5 m/s and the one that features a transition in the dust fragmentation velocity is small, even at the ice line.}
\end{figure}

\end{appendix}

\end{document}